\documentclass[preprints,review,accept,moreauthors,pdftex]{Definitions/mdpi}
\usepackage{comment} % ADDED BY MM

\firstpage{1}
\pubvolume{xx}
\issuenum{1}
\articlenumber{5}
\pubyear{2021}
\copyrightyear{2020}
\history{\tiny{Received: December 1, 2020}}
\Title{A Review of Basic Energy Reconstruction Techniques in~Liquid Xenon and Argon Detectors for Dark Matter and~Neutrino Physics Using NEST}

\Author{Matthew Szydagis $^{1,}$*\orcidA{}, Grant A.~Block $^{1,2}$, Collin Farquhar $^{1,3,4}$\orcidB{}, Alexander J.~Flesher $^{{5}}$\orcidH{}, Ekaterina~S.~Kozlova $^{6,7}$\orcidG{}, Cecilia Levy $^{1}$\orcidC{}, Emily A.~Mangus $^{1,8}$, Michael Mooney $^{5}$\orcidD{}, Justin Mueller $^{5}$\orcidE{}, Gregory R.C.~Rischbieter $^{1}$\orcidF{}, and  Andrew K.~Schwartz $^{1,9}$}

\address{\tiny{$^{1}$ \quad Department of Physics, University at Albany SUNY, 1400 Washington Ave., Albany, NY 12222-0100, USA \\
$^{2}$ \quad Department of Physics, Applied Physics and Astronomy, Rensselaer Polytechnic Institute, Troy, NY 12180, USA \\
$^{3}$ \quad The College of St.~Rose, Albany, NY 12203, USA  \\
$^{4}$ \quad ANDRO Computational Solutions, LLC, Rome, NY 13440, USA \\
$^{5}$ \quad Department of Physics, Colorado State University, Fort Collins, CO 80523, USA \\
$^{6}$ \quad Institute for Theoretical and Experimental Physics Named by A.I.~Alikhanov of National Research Centre “Kurchatov~Institute”, 117218 Moscow, Russia \\
$^{7}$ \quad Moscow Engineering Physics Institute (MEPhI), National Research Nuclear University 115409 Moscow,  Russia \\ $^{8}$ \quad Northrop Grumman, Goddard Space Flight Center, Greenbelt, MD 20771, USA \\
$^{9}$ \quad Department of Physics and Astronomy, Rutgers University, Piscataway, NJ 08854, USA}}

\corres{\tiny{Correspondence: mszydagis@albany.edu}\vspace{-7pt}}

\abstract{Detectors based upon the noble elements, especially liquid xenon as well as liquid argon, as both single- and dual-phase types, require reconstruction of the energies of interacting particles, both in the field of direct detection of dark matter (weakly interacting massive particles WIMPs, axions, etc.) and in neutrino physics. Experimentalists, as well as theorists who reanalyze/reinterpret experimental data, have used a few different techniques over the past few decades. In this paper, we review techniques based on solely the primary scintillation channel, the ionization or secondary channel available at non-zero drift electric fields, and combined techniques that include a simple linear combination and weighted averages, with a brief discussion of the application of profile likelihood, maximum likelihood, and machine learning. Comparing results for electron recoils (beta and gamma interactions) and nuclear recoils (primarily from neutrons) from the Noble Element Simulation Technique (NEST) simulation to available data, we confirm that combining all available information generates higher-precision means, lower widths (energy resolution), and more symmetric shapes (approximately Gaussian) especially at keV-scale energies, with the symmetry even greater when thresholding is addressed. Near thresholds, bias from upward fluctuations matters. For MeV-GeV scales, if only one channel is utilized, an ionization-only-based energy scale outperforms scintillation; channel combination remains beneficial. We discuss here what major collaborations use.}


\begin{document}

%%%%%%%%%%%%%%%%%%%%%%%%%%%%%%%%%%%%%%%%%%

%%%%%%%%%%%%%%%%%%%%%%%%%%%%%%%%%%%%%%%%%%
\vspace{-15pt}
\section{Introduction}
\vspace{-5pt}
The noble elements, especially xenon (Xe) and argon (Ar) as liquids, have been instrumental in the field of dark matter (DM) direct detection, focused on identifying the missing $\sim$25\% of the mass--energy content of the universe. They have also been key for neutrinos. In the former case, Xe~\cite{Akerib_2020_LZ_ProjSens,Aprile_2017_Xe1T1st,PhysRevLett.119.181302} and Ar~\cite{Li_2019_DS50_1st,Ajaj_2019} are each used by distinct large collaborations, and used both to search for continuous spectra, such as the approximate falling exponential expected from the traditional weakly interacting massive particle (WIMP)~\cite{McCabe_2010}, or monoenergetic peaks expected from dark photons or bosonic super-WIMPs~\cite{Aprile_2017_Boson}. In the latter case, argon is used in long- and short-baseline oscillation studies~\cite{abi2020deep,microboonecollaboration2020convolutional} and xenon in the search for neutrinoless double-beta decay, as either a liquid~\cite{Anton_2019_EXO-200_Complete} or a gas~\cite{Woodruff_2020}. In all of these cases, there is a clear need for high accuracy and high precision in energy reconstruction and good energy resolution in order to identify signals and backgrounds, and calibrate the detectors. Combining the data with high-fidelity Monte Carlo (MC) simulations can aid in this task. Xe and Ar produce scintillation light, and when an external electric field is applied then ionization electrons can be extracted as well. In a dual-phase time projection chamber (TPC), a gas stage converts the ionization into a secondary scintillation pulse~\cite{Akerib_2013_NIM_Det} while a single-phase TPC reads out the charge directly~\cite{2014_EXO-200_First,Fernandes_2010}. Energy scales have been based in the past and present on the scintillation, on the ionization, and on their combination.

In this work, we will be reviewing each of these methods, contrasting them and enumerating their strengths and weaknesses, in terms of the mean, median, and mode (e.g., Gaussian peak centroid) of reconstructed energy best matching the true energy (MC truth energies and/or monoenergetic calibration peaks), the width, and the shape (symmetry). Multiple types of particles will be covered, addressing scattering from atomic electrons and nuclei, electron recoil (ER) and nuclear recoil (NR) respectively, and recoil energies from sub-keV up to GeV, from DM-WIMP-induced NR or coherent elastic neutrino nucleus scattering (CE$\nu$NS), to neutrino-induced ER. The summaries in each section make idealized recommendations, with detector-dependent caveats, for future DM/neutrino~projects.

%%%%%%%%%%%%%%%%%%%%%%%%%%%%%%%%%%%%%%%%%%
\section{Methods}
\vspace{-15pt}
\subsection{General}

Examples of usages of each of the possible energy scale definitions are taken from empirical data wherever possible, but also compared to the Noble Element Simulation Technique (NEST), which is also used by itself where data are lacking. NEST is a global, experiment- and detector-independent MC framework that allows simulation of scintillation and ionization yield averages and resolutions as functions of incoming or deposited energy, electric field, and interaction type~\cite{NESTWebSite}.

The values of detector-specific parameters also need to be known in order to permit NEST to simulate the detectors effects for a specific experiment. The most important numbers are $g_1$, $g_2$, and the magnitudes of the drift and extraction electric fields, if applicable. $g_1$ and $g_2$ are respectively defined as the gains of the primary and secondary scintillation channels, the latter from ionization (again, only if applicable). $g_1$ is always between 0 and 1, and is an efficiency which combines the quantum efficiencies of one's photo-sensors with the geometric light collection efficiency (it is also known as the photon detection efficiency). It can include or exclude, depending on choice of units, the probability for certain photon detectors (especially in the vacuum ultraviolet or VUV) to produce more than one photoelectron (phe) for a single incoming photon~\cite{Faham_2015}. Typical values across all experiments using Xe or Ar are $\sim$0.05--0.20 phe per photon~\cite{Aprile_2010_Xe1001st,Akerib_2016_Run03ReAnal,Akerib_2017_Run04,Abe_2014,Hackett:2017jnd,Agnes:2016mgq,Pagani:2017}. $g_2$ is a combination of the electron extraction efficiency for a two-phase TPC with the gas gain, i.e., the number of photons produced per extracted electron times the photon detection efficiency in the gas phase~\cite{Akerib_2020_GregRC14}. A typical value is $\sim$10--30 phe per electron.

Some photons, especially in the ultraviolet range, carry sufficient energy in order to generate more than one phe per incident photon, in the photocathodes of certain photon detectors, and stochastically, not consistently. The definition of the unit ``photons detected'' or ``detected photons'' or phd for short is simply phe (alternatively called PE by some authors) divided by (1 + $p_{2e}$), where $p_{2e}$ is the probability of producing two (photo)electrons, typically 0.1--0.3, depending on the manufacturer, temperature, and individual phototube: $phd = phe/( 1 + p_{2e} )$, while a similar translation applies to the $g_1$. While taking this effect into account is ideal for the resolving of peaks and achieving the best possible background discrimination for one's final analyses, the convention for how the final results are plotted varies by experimental collaboration, making comparisons more difficult (the XENON and DARWIN groups prefer phe or PE but LUX and LZ prefer phd). PIXeY's 2-phe probability was reported as 17.5\%~\cite{boulton}, implying that division by 1.175 would convert phe into phd.

We define $N_{ph}$ and $N_{e-}$, as the original numbers of photons and electrons generated from an interaction site, determined by MC truth (integer) or empirical reconstruction (float). $N_q$ is their sum. In data, or advanced MC, including detector effects like finite detection efficiencies not just mean yields: $N_{ph} = S1_{c}/g_{1}$ and $N_{e-} = S2_{c}/g_{2}$. S1 and S2 are the primary and secondary scintillation pulse areas. Subscript ‘c' denotes correction, primarily for XYZ-position effects, as light collection efficiency may depend significantly on 3D position, especially in a large-scale detector~\cite{Akerib_20173dKr83}. The energy dependence is intrinsic to the element, unrelated to the position inside a particular detector. It is thus useful to also define two additional terms, $L_y$ and $Q_y$, which respectively refer to the $N_{ph}$ and $N_{e-}$ per unit of energy.

\textls[+15]Our default guiding formula, at least for combined energy-scale reconstruction, is therefore: 

\begin{equation}
E = N_q \frac{W_q}{L(E)} =  (N_{ph} + N_{e-} ) \frac{W_q}{L(E)} = ( \frac{S1_{c}}{g_{1}} + \frac{S2_{c}}{g_{2}} ) \frac{W_q}{L(E)}.
\label{Eqn1}
\end{equation}

$L$ is Lindhard factor, quantifying $E$ ``lost'' (if a detector cannot see phonons) into heat (atomic motion) instead of observable quanta. Called also ``quenching'' historically, that word is not precise: quanta are not being quenched {per~se} but are not being created {ab initio}, unlike in quenching by impurities or ionization density. It depends on $E$, making Equation~(1) circular. Circularity can be avoided by a good MC model like NEST, and by quasi-monoenergetic NR data, as in the LUX D-D analysis~\cite{luxcollaboration2016lowenergy}. For ER, $L$ is taken to be 1.0, not implying no heat loss but an approximately constant loss as a function of $E$ that can be ``rolled into'' the definition of $W_q$, raising it. For neutrino experiments, $E$ depositions are tracks not point-like, and $dE/dx$ (energy loss per unit of distance) is more~relevant.

The $L$ quantifies the effectiveness of an initial NR at producing more elastic recoils as secondary interactions not inelastic, i.e., atomic excitations/ionizations. It can be modeled in other/better ways than Lindhard's, expected to break down at low $E$s, as seen in Si/Ge~\cite{Sarkis_2020}. It should not be confused with $L_{eff}$ used for S1~\cite{Angle_2008_1stXe10,Manzur:2009hp,Plante_2011,Aprile_2010_Xe1001st} at zero field, and not accounting for $^{57}$Co 122~keV $\gamma$ rays not being representative of even ER yields, at different $E$s~\cite{Szydagis_2011,Aprile_2012_Compton} ($L_{eff}$ was the ratio of NR to ER light yields).

$L(E)$ can be thought of as merging the scintillation and ionization efficiencies (but $L_{eff}$ is only scintillation, again compared to ER $L_y$). However, as with ``quenching,'' it is better to avoid ``efficiency,'' which can be confused with the DAQ and analysis, where a certain number of photomultiplier tubes (PMTs) must fire to count an S1 (2-fold coincidence for example in LUX, a much later example using ER, where this coincidence requirement, $g_1$ $\ll$ 1, and other effects lead to a threshold). The $E$ within the $L$-factor is the true energy of a nucleus recoiling from a neutron (or $\nu$ coherently, or DM hopefully in the future). When quoting imperfect reconstructed energy using Equation~(1), with $0 < L < 1$, the standard unit is keV$_{nr}$ to contrast with the unit defined assuming $L = 1$ for ER earlier: keV$_{ee}$.

The subscript `q' indicates that the $W$ value is not based on scintillation or ionization alone. $W_q$ is effectively an average over microphysical processes producing excited/ionized atoms. We do not describe them, as they are within other works~\cite{Szydagis_2013,Sorensen_2011,Dahl:2009nta}, nor establish NEST's accuracy here, using it for convenience where data are lacking. The physical details and how NEST captures them are beyond the scope of this paper. In this section, in particular, we summarize established methods~\cite{Szydagis_2011,szydagis2014detailed}.

At zero field, one only has access to the S1, and for specialized searches for (sub-)GeV DM~\cite{Angle_2011,Aprile_2019_S2Only} to only S2 due to the low energies involved, as the ionization, i.e., charge yield is typically larger and easier to detect, even as $E$ goes to zero. The formula correspondingly morphs into one of these:

\begin{center}
$ E = \frac{N_{ph}}{L_y} = \frac{S1_{c}}{g_{1}}\frac{1}{L_y(E, \mathcal{E})} \quad (2) \quad OR \quad E = \frac{N_{e-}}{Q_y} = \frac{S2_{c}}{g_{2}}\frac{1}{Q_y(E, \mathcal{E})}$, \quad (3)
\label{Eqn2a3}
\end{center}

where $L_y = N_{ph} / E$ and $Q_y = N_{e-} / E$ are functions of energy $E$ and field $\mathcal{E}$. Each differs~for~ER~and NR; they are not fixed as $W_q$ was, for ER. The challenge of $E$ reconstruction increases via an inherent non-linearity: e.g., 2x the $E$ does not mean 2x the light or charge as it does with $N_q$ at least for ER. When considering resolution, this causes deviation from the Poisson expectation of $1/\sqrt{E}$ improvement~with higher $E$, which only applies for the combined scale ($1/\sqrt{N_q}$). A more general power law then~works best~\cite{Aprile_2011_W14,Aprile_2020_resE,Aprile_2020_Excess}. In liquid xenon (LXe), the fact $L_y$ is not flat vs.~$E$ was first demonstrated by Obodovskii and Ospanov~\cite{OboAndOsp_1994} for ER (implying for fixed $W$ that ER $Q_y$ is also not constant) and by Aprile et~al.~for NR $Q_y$~\cite{PhysRevD.72.072006,PhysRevLett.97.081302} but not well known within the DM field for ER background at least until much later publications~\cite{Aprile_2012_Compton,Baudis_2013}. In the case of experiments like EXO (0$\nu\beta\beta$ decay) and DUNE (long-baseline $\nu$'s) using Xe/Ar, only the first half of (3) applies, as they measure $Q$ directly in liquid using wire readout planes, instead of in gas. In addition, for combined $E$ (Equation~(1)) especially, S2 can be defined using a subset of photon sensors instead of all. A cylindrical TPC typically has two arrays, one at each end. The bottom one only (subscript `b' for differentiation from total S2) may be used for $E$ reconstruction as in \cite{Aprile_2018_OneTonYear} to adjust for light loss created by inoperative tubes in gas, or saturation. Lastly, $g_1$ can be vastly smaller in kilotonne-scale $\nu$ experiments, unlike the range quoted earlier for DM: they rely on $Q_y$.

For the purposes of reproducibility, we note that all of the work presented here uses the latest stable NEST release at time of writing: Version 2.2.0~\cite{szydagis_m_2018_4062516} for which the default detector parameter file is designed to mimic the first science run of LUX~\cite{Akerib_2014} but which we modify as needed to reproduce other experiments, focusing again on $g_1$, $g_2$, and drift electric field $\mathcal{E}$ as the three most salient inputs.

\subsection{NEST-Specific Details}

The main assumptions utilized in NEST to reproduce efficiencies are briefly explained~below.

\vspace{-5pt}
\begin{enumerate}
\small
\item A Fano-like factor sets variation in total quanta, with a binomial distribution for differentiating excitons/ions (inelastic scattering). In LXe, it is not sub-Poissonian as in GXe, as experimentally verified in each phase~\cite{YAHLALI2010520}.
\item Recombination fluctuations~\cite{Conti_2003,Akerib_2017_EvanP,luxcollaboration2020discrimination}: the ``slosh'' of $N_{ph}$ vs.~$N_{e-}$ caused by recombination probability for ionization $e^-$s, which may either recombine to make more S1, or escape to make S2. These are worse (i.e., larger) than expected na\"ively (non-binomial). This is distinct from Fano factor, and canceled by combined $E$.
\normalsize
\end{enumerate}
\vspace{-5pt}

The above are general, but the below depend on the detector, all combining into the final efficiency.

\vspace{-5pt}
\begin{enumerate}
\small
\item $g_1$ is used to define a binomial distribution for the S1 photon detection efficiency with $<S1> = g_1 N_{ph}$.
\item For an S1 to be above the trigger threshold, most experiments require that $O$(0.1)~phe must be observed in $N$ PMTs for $N$-fold coincidence, where usually $N = 2$ or 3, within a coincidence window, of 50--150~ns, requiring a basic timing model for singlet and triplet states and photon propagation time. The 2 or 3-fold coincidence prevents triggering on photo-sensor dark counts. Baseline noise $O$(0.1)~phe is also modeled. 
\item The pulse areas of single phe are assumed to follow a truncated (negative phe are not possible) Gaussian distribution, with $O$(10\%) resolution differing by photon-sensor, but a detector-wide average is used for NEST, as an approximation. If single phe detection efficiency is reported, it can be used instead of a threshold applied to a Gaussian random number generator, thus taking non-Gaussianity and other detector-specific idiosyncrasies into account. This and others numbers are collected from arXiv, publications, and theses.
\item Drifting, diffusing electrons are removed via an exponential electron lifetime, and are assumed to follow a binomial extraction efficiency, while the number of photons produced per surviving extracted electron depends on the gas density, electroluminescence electric field, and gas gap size, in a 2-phase TPC~\cite{Mock_2014,Chepel_2013}.
\item A special Fano factor, typically also >1 for S2 accounts for non-Poissonian behavior, due for example to grid wire sagging. A value of 2--4 is normal~\cite{araujo2020revised}. S2 photons experience a similar binomial photon detection efficiency as S1 photons, moving along from photons to phe (for S2, from electrons to photons to phe). A raw, total S2 threshold $O$(100)~phe removes the lowest-energy events, to avoid few-electron backgrounds~\cite{akerib2020investigation}.
\item S1 and S2 XYZ variation is simulated in NEST if provided in analytical form, then realistically corrected back out, based upon finite position resolution, not MC truth positions, thus allowing not only for correct means but correct widths. (Z or drift correction applies only to S1, handled for S2 by the electron lifetime.)
\item $N_{ph}$ falls while $N_{e-}$ rises with drift field in anti-correlated fashion, and fields can be non-uniform.
\normalsize
\end{enumerate}
\vspace{-0pt}

A final step of noise is applied as an empirical smearing to the S1 and S2 pulse areas to match realistic experimental data; however, the above lists capture the vast majority of fluctuations that can shift the low-energy efficiency higher or lower. Additional noise is typically at a level of $O$(1\%) from unknown sources, but likely due to position correction imperfection and other analysis-specific effects, discussed in \cite{Dahl:2009nta,Akerib_2020_GregRC14,szydagis2020investigating}. This is uncorrelated noise, as it is applied separately to S1 and S2, while the variation induced by the Fano factor (Step 1 in first list) is correlated ``noise'' due to raising S1 and S2 together, and the recombination fluctuations constitute anti-correlated noise, as in raising S1 they lower S2, and vice versa. All fluctuations can move events above and/or below nominal thresholds.

All steps above are uncertain especially when it comes to a simulation software package such as NEST. It would be infeasible to discuss and address all uncertainties. Thus, we will mention only one that is often largest. The $N_{ex}$ and $N_i$, followed by $N_{ph}$ and $N_{e-}$, produced at Step (1.) in the first list above, depend on particle, energy, field, density via temperature and pressure, and phase, and at the lowest energies (sub-keV especially, and in particular for NRs) there is a non-negligible systematic uncertainty from the values assumed for the average yields ($L_{y}$ and $Q_{y}$) which are always the first step in NEST modeling. For ER $Q_y$, the discrepancy between different data sets and models is as much as 20\% below 1~keV, as illustrated by contrasting NEST~\cite{szydagis2020investigating} with PIXeY $^{37}$Ar data~\cite{Boulton_2017}, X-rays in LUX~\cite{Akerib_2017_DQ}, and $^3$H in XENON100~\cite{Aprile_2018_Discrim}. $L_y=0$ in NEST for sub-keV ER but not in XENON models, which thus predict higher efficiency~\cite{Aprile_2020_Excess}. The 50\% efficiency point typically is the ``$E$ threshold.''

For the purposes of this review paper, what is of greatest importance is that the assumptions, of the default NEST yield models, are not varied, when comparing different energy reconstruction techniques, so that at the very least a robust comparison can be made among them. That being said, we stress the accuracy of NEST in reproducing efficiency even for publications where the authors have no access to the original data is high (not only LUX~\cite{Akerib_2018_PRD}). XENON1T is an excellent example~\cite{szydagis2020investigating}.

%%%%%%%%%%%%%%%%%%%%%%%%%%%%%%%%%%%%%%%%%%
\section{Results}

Xe is examined first (ER then NR) followed by Ar. For Xe, the present-day relevant experiments seeking DM for whom this review is most pertinent include: LZ, XENON, and PandaX. The use of LAr is divided, present and past, across DEAP, CLEAN, ArDM, and DarkSide (first two single-phase and zero field, latter two dual-phase and non-zero electric field) on the DM front, and DUNE, MicroBooNE, ArgoNeuT, ICARUS, plus many others, studying neutrinos. Enriched LXe is used by nEXO, a TPC but only one phase, and NEXT (GXe) for the hunt for 0$\nu\beta\beta$ decays. $\alpha$s and heavier ions different from the medium, with properties like additional quenching, modify the $E$ reconstruction formulae, but we will only focus on basic ER and NR; other recoil types are already covered elsewhere~\cite{NESTWebSite,CutterTalk}.

%%%%%%%%%%%%%%%%%%%%%%%%%%%%%%%%%%%%%%%%%%
\subsection{Liquid Xenon Electron Recoil}
\unskip
\vspace{-15pt}
\subsubsection{\textbf{Low Energy: keV-scale (Dark Matter Background, Signal) Basic Recon of Mono-\textit{E}~Peaks}}

The term electron recoil or ER refers to interactions with the electron cloud, such as from beta emissions and the Compton scattering or photoabsorption of gamma rays. In a WIMP search, ER is the primary background, but in a more general DM (or exotic physics) search a monoenergetic peak or even continuous ER spectrum can be the signal~\cite{Akerib_2017_AxionALP}. To illustrate the differences among the S1-only, S2-only or Q-only, and combined-$E$ scales for reconstruction the first example is the lowest-energy ER calibration peak available at time of writing for LXe where we have S1 and S2 still: the electron capture decay of $^{37}$Ar at 2.82~keV. It is also a timely example due to the recent XENON1T ER result~\cite{Aprile_2020_Excess,szydagis2020investigating}.

\begin{figure}[p!]
\includegraphics[width=0.9\textwidth,clip]{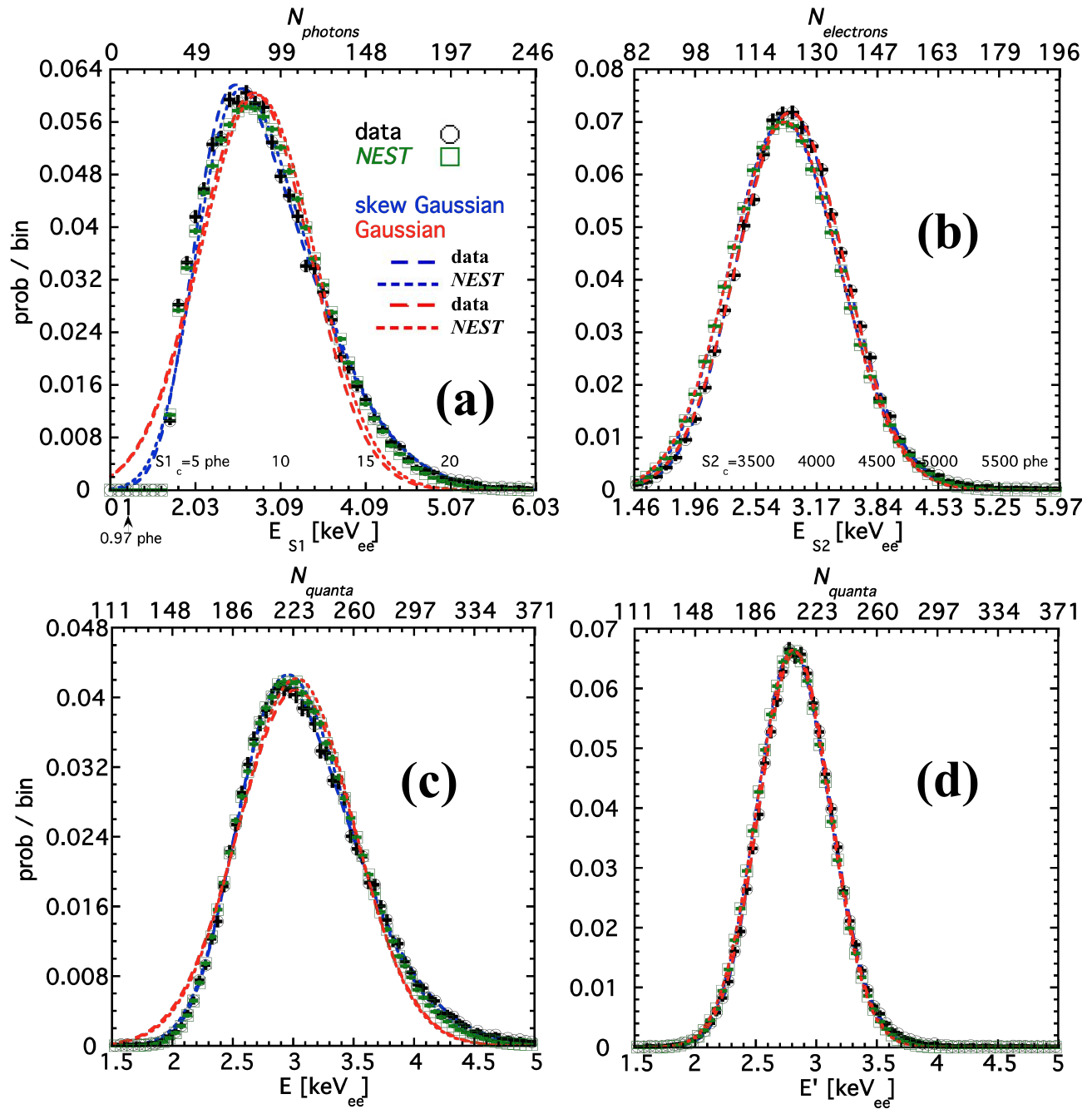}
\vspace{-0pt}
\caption{Reconstruction of the 2.8224~keV $^{37}$Ar peak in the PIXeY detector~\cite{Boulton_2017} compared to the Noble Element Simulation Technique (NEST). Real data are always hollow black circles, NEST Monte Carlo (MC) data are green squares. Gaussian fits are in red, skew Gaussian (better fit) in blue, with the fits to data in long dash and NEST in short dash (indistinguishable due to NEST's fidelity). Number of events in data $7.4 \times 10^4$, while $9.3 \times 10^5$ in the MC, after all cuts (i.e., all thresholds). (\textbf{a})~Original, non-linear S1-only $E$ scale used for liquid xenon (LXe). Bins with non-zero counts begin very suddenly at the left in both NEST and data due to a cut-off created by triggering only on 3-fold PMT (Photo-Multiplier Tube) coincidence, and the other threshold requirements. The results are highly skewed, driving the asymmetry within the combined-energy fit later. (\textbf{b}) S2-only, which is quite symmetric, so Gaussian and skew-Gaussian fits overlap. (\textbf{c}) The combined-$E$ scale in common use now for LXe dark matter (DM) detectors. Gaussian fits in red are clearly poorer compared to skew fits, diverging from the histogram in the cases of NEST and data alike. (\textbf{d}) An optimized combination for energy, as done on PIXeY. Both NEST and data, and Gaussian and skew-Gaussian fits alike, have all become indistinguishable for this stage. The best-fit mean energy has shifted from 3.03~keV in (\textbf{c}) to 2.82~keV for (\textbf{d}). This improvement in precision is also reflected in the sum of the mean quanta from (\textbf{a}) and (\textbf{b}) matching (\textbf{d}), but not (\textbf{c}), which is too high. The skew parameter $\alpha$ decreases from 3 for S1 only (\textbf{a}) to 2 for the combined scale in (\textbf{c}) and 1 in (\textbf{b}, \textbf{d}).}
\label{Fig1}
\vspace{-0pt}
\end{figure}

While this $E$ is not often in the regime of efficiency drop, we do address it in this section. Our source of data here was the seminal work led by McKinsey et al.~at Yale/Berkeley~\cite{Boulton_2017} who constructed a small-scale calibration chamber, PIXeY, with $g_1 = 0.097 \pm 0.007$ phe/photon and $g_2 = (0.78 \pm 0.05) \times 36.88 = 28.77 \pm 1.85$~phe/$e^-$ (extraction efficiency times the single-$e^-$ pulse area). To replicate PIXeY precisely with NEST in Figure~\ref{Fig1}, we use $g_1 = 0.1015, g_2 = 30.65$, only 0.6--1.0$\sigma$ higher. We  studied only 198~V/cm. $W_q$ was taken to be 13.5~eV, between Dahl~\cite{Dahl:2009nta} and neriX~\cite{Goetzke_2017} measurements. In all plots, keV$_{ee}$ means reconstructed energy ($ee$ standing for electron equivalent, for when NR is translated into this scale) as opposed to MC truth, or a known single $E$ from a peak, reported in keV sans subscript.

While lower-$E$ calibrations exist than 2.8~keV, this is the lowest where S1 and S2 are identified separately. Others become S2-only~\cite{Akerib_2017_DQ}. Figure~\ref{Fig1}a demonstrates that the S1-only scale, used primarily on XENON10~\cite{Angle_2008_1stXe10} and continuing on in a subset of XENON100 analyses~\cite{Aprile_2014_Axion}, performs the poorest, with an energy resolution $\sigma/\mu$ of 38.63\% for data (38.59~NEST) in red. (Both XENON10/100 had similar $g_1$s to PIXeY, but slightly lower, at $\sim$0.07.) Values in parentheses following values from data now are for NEST, on to subsequent pages, in the style of A (B). An S1 basis is further complicated by non-linearity. Next, in Figure~\ref{Fig1}b the S2 only is plotted, leading to an 11.65\% (12.20) resolution. While we see later that the combined scale, by including all available information, is typically best at higher energies, this is not the case any longer at keV-scale energies, as (c) indicates. Combined resolution is 15.84\% (15.51). This is caused by data from these $E$s being comprised of upward S1 fluctuations above nominal S1 threshold, due to finite $g_1$. An experimenter measures the right tail of S1s essentially.

Nevertheless, it is possible to mitigate S1 effects, still include S1 in the $E$ calculation, and obtain the best possible resolution. One way is to fit a skew Gaussian (parameters explained in \cite{luxcollaboration2020discrimination,szydagis2020investigating}):

\addtocounter{equation}{2}
\vspace{-0pt}
\begin{equation}
y = A e^{\frac{-(x-\xi)^{2}}{2\omega^{2}}} [ 1 + erf ( \alpha \frac{x-\xi}{\omega \sqrt{2}} ) ]
\label{Eqn4}
\end{equation}

While this has been done for bins in S2 vs.~S1~\cite{luxcollaboration2020discrimination} and once for combined $E$~\cite{szydagis2020investigating}, it is most effective for S1: see the improvement in Figure~\ref{Fig1}a (blue vs.~red). A skew fit, while including an error function $erf$ and similar to the equation used in \cite{Boulton_2017} that should account for triggering, still misses some points, as keV-level S1 becomes non-Gaussian and non-symmetric due to trigger efficiency dropping below 100\%. In plot (c) however, the reduced $\chi^2$ drops from $O$(100) for both data and NEST to 2.6 for the 1.5--5~keV$_{ee}$ range, still too high due to features in the data not captured even by a skew-normal fit, but more sensible. Asymmetries arise from both thresholding bias~\cite{luxcollaboration2016lowenergy} and microphysics~\cite{szydagis2020investigating,luxcollaboration2020discrimination}.

\subsubsection{\textbf{More Advanced Energy Reconstruction Strategies, from keV to MeV Scales, and~Resolution}}

A superior mitigation strategy can be found upon realization that the optimal weights for the S1 and S2 pulse areas are no longer simply $g_1$ and $g_2$ at the $O$(keV) scale. We can recast this statement in terms of the (n)EXO-style combined-energy scale first developed by Conti~\cite{Conti_2003}: instead of using a $g_1$ and $g_2$, it defines what is known as an angle of anti-correlation for summing S1 plus Q or S2. As energy decreases, the angle becomes energy-dependent instead of being fixed as $tan^{-1}(g_2/g_1)$~\cite{Aprile_2011_W14} and thus no longer respecting ``perfect'' anti-correlation of quanta, with $N_{ph}$ and $N_{e-}$ always summing to $N_q=E/W_q$. Note there is no evidence of anti-correlation breakdown at least in LXe above 1~keV: this effect is caused by inability to reconstruct $N_{ph}$ well in data due to dropping S1 efficiency, as first suggested by Szydagis (2012) and first publicly applied in the PIXeY $^{37}$Ar paper~\cite{Boulton_2017}.

\begin{equation}
E' = ( w_1(E) \frac{S1_{c}}{g_{1}} + \frac{S2_{c}}{g_{2}} ) W_q  \times w_2(w_1)
\label{Eqn5}
\end{equation}

Parameter $w_1$ decreases the S1 weight for low $E$s, countering thresholding; one could increase S2 weight, but this is equivalent. Multiplying S1 by one weight, and S2 by another would be redundant. Instead, one is applied to S1, and a second $w_2$ to the formula as a whole to bring the average of the $E$ being reconstructed back to the correct mean after the shift caused by adding $w_1$, while simultaneously correcting for any efficiency bias near the S1 and/or S2 thresholds. It is not technically independent then, thus written in Equation~(5) as a function of $w_1$, which itself is a function of energy. To avoid a circularity, Equation~(1) can be used to determine its $E$ dependence, for use in (5), and the process can be iterative, defining E'' after E', etc. Knowledge of the proper weights \textit{a~priori} is achievable via MC.

Without an MC such as NEST tuned using earlier calibration data, it is possible to empirically determine the two weights by calibrating an experiment with monoenergetic peaks ($e^-$ capture, X-ray, gamma-ray). In the case of PIXeY's $^{37}$Ar measurement, the values which minimize the width of the Ar peak in reconstructed energy (optimum resolution) are $w_1 = 0.19$ and $w_2 = 1.38$ (NEST: 0.23, 1.35). Looking back at Figure~\ref{Fig1}d, the very positive effect of applying Equation~(5) is evident: the resolution is 10.60\% (10.68\%) close to S2-only~(b) but a factor of $\sim$1.5 improvement over (c) the more traditional ``plain'' combined scale. Moreover, the Gaussian centroid has dropped from 3.03~keV$_{ee}$ (again, higher because of triggering on high-S1 fluctuations) to 2.83 (2.81) much closer to the true value of 2.82, while asymmetry in the histogram has nearly vanished, with the best-fit skew Gaussian possessing a positive (right) skewness parameter $\alpha$ $<$ 1 (compared to $\sim$2--2.5 in (c), which used Equation~(1)). While it is infeasible to remove all skew, as some is intrinsic (from recombination physics)~\cite{luxcollaboration2020discrimination}, in panel (d) the Gaussian (red) and skew (blue) fits are nearly indistinguishable (unlike in plot 1c).

This technique, essentially equivalent to inverse-variance weighting, is not widespread for DM, although found in \cite{boulton}. From its inception, LUX has relied on Equation~(1), i.e., plot 1c's method, but a  Profile Likelihood Ratio (PLR) analysis effectively takes into account energy bias by relying on NEST to produce non-analytic 2D PDFs for both background and signal, relying on MC truth energy converted to S1 and S2, not reconstructed energy~\cite{Akerib_2012_LUXSim}. XENON's own MCs for its PLR perform the same function, taking MC truth and/or calibrations as input, and outputting (S1, S2) PDFs mimicking data~\cite{Aprile_2019_Xe1TmcPLR}.

Many possible enhancements exist, like Maximum Likelihood~\cite{bloch2020exploring} and Machine Learning~\cite{Delaquis_2018}. These can take more than S1 and S2 into account, e.g., bypassing calibrated 3D corrections, feeding raw S1 and S2 plus positions into an artificial neural network (ANN) or Boosted Decision Tree (BDT) from which XYZ dependence emerges, given sufficient training. Ernst and Carrara suggest that it is possible to determine how much is sufficient~\cite{carrara2017upper}. Some examples of additional training variables include the $E$-dependent S1 pulse shape, usable given sufficient statistics~\cite{Mock_2014,Akerib_2018_PSD}, as well as breakdown of pulse areas into top and bottom arrays, capitalizing on anti-correlation of top~vs.~bottom light similar to that of S1~vs.~S2, as used early on LUX~\cite{Akerib_2013_Run02}, good for detectors sans S2 (0~V/cm). However, idiosyncrasies of individual detectors make it difficult to review such methods, which still rely on S1 and S2 as the two most important variables regardless. ANNs/BDTs are best trained by a combination of S1s and S2s from data and MC, per analysis, and have so far never been applied at <1 MeV.

\begin{figure}[ht]
\includegraphics[width=0.99\textwidth,clip]{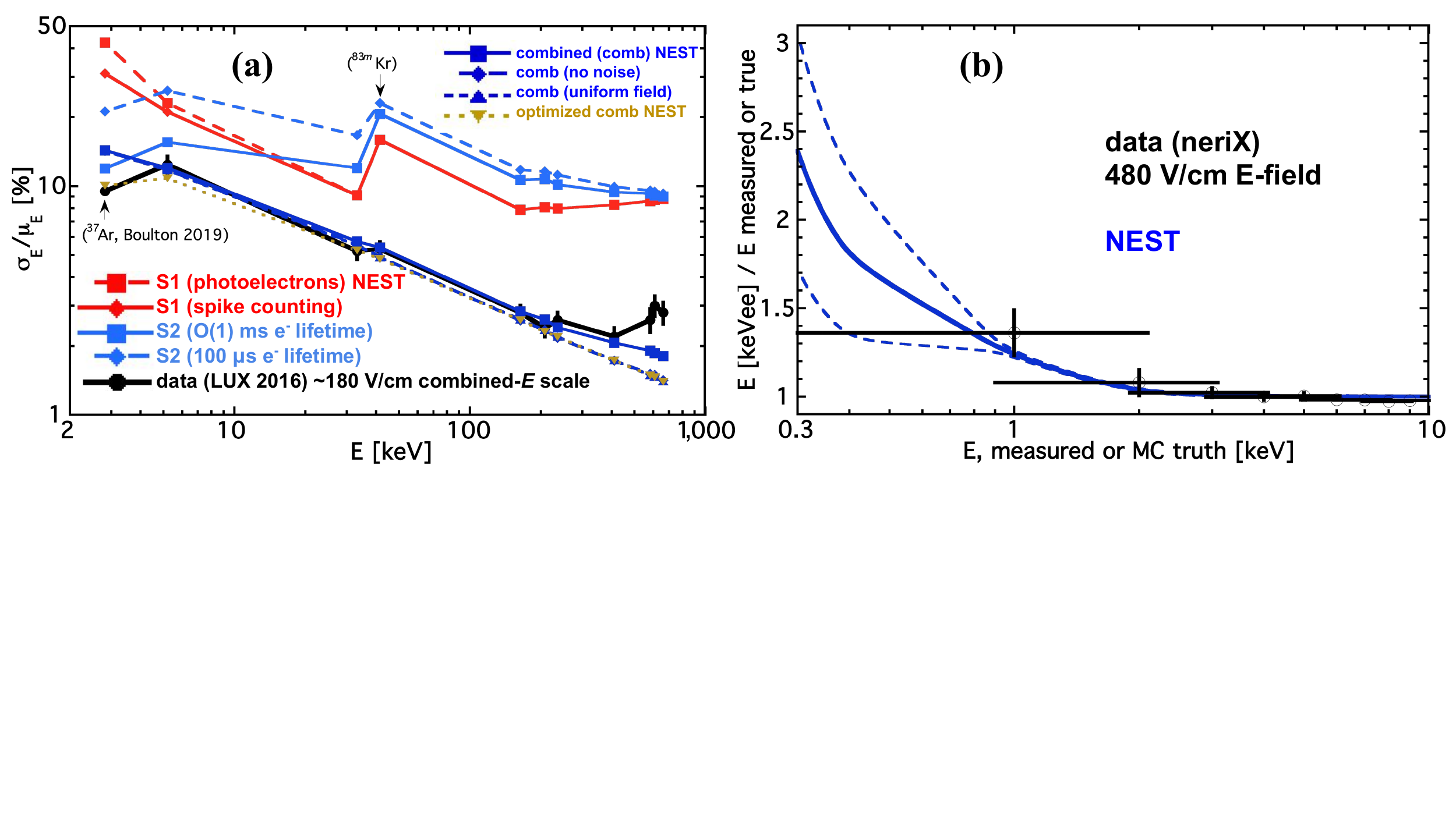}
\vspace{-111pt}
\caption{Review of $E$ resolutions and means. (\textbf{a}) Resolution vs.~$E$ for LUX monoenergetic calibration plus background~\cite{Akerib_2017_EvanP}. Only combined resolutions published (black), but S1 (red) and S2 (cyan) scales included from NEST, to show they are mutually comparable at $O$(10--1000) keV. The lowest-$E$ point is optimized as done in PIXeY, so anomalously good (low). NEST combined scale blue (below S2 only in cyan except for $^{37}$Ar) and optimized gold. $w_1$ in optimal scale varies from 0.26 to 1.0 from 2.8 to 662~keV, while $w_2$ falls from 1.45 to 1.00. (Lines are guides not fits.) (\textbf{b}) Data from neriX (Columbia's small-scale calibration chamber, like PIXeY) as black points vs.~measured recoil energies from Compton scatters~\cite{Goetzke_2017} compared to NEST in blue vs.~true energy known from MC, showing consistent deviation for both in reconstructed energies for a combined but non-optimized scale due to threshold bias (weights as used for NEST in gold in (\textbf{a}) correcting for this not applied purposely in blue in (\textbf{b}) to show this effect).}
\vspace{-6pt}
\label{Fig2}
\end{figure}

Given these difficulties, this paper focuses only on the energy reconstruction scales which can be formulated purely analytically with relative ease: again, S1-only (Equation~(2)), S2-only (Equation~(3)), combined (Equation~(1)), and the so-called optimized scale $E'$ (Equation~(5)). To broaden applicability to more detectors, we also consider variants. Figure~\ref{Fig2}a shows S1-only in red and S2-only in cyan. The dashed red line illustrates how the S1 scale is poorer (the effect propagates into combined energy) when one does not account for the so-called ``2-phe effect,'' mentioned earlier~\cite{Faham_2015}. Accounting for this via dividing it out improves the resolution, as the additional phe do not provide any new information on the original number of photons produced $N_{ph}$, even though they may be useful in lowering threshold and increasing the sensitivity to lower-mass DM~\cite{Akerib_2020_Nellie}. The solid red NEST line demonstrates what improvement is achieved in doing this, plus attempting to reconstruct the integer numbers of photons hitting the photomultiplier tubes (PMTs), instead of only reporting S1 pulse areas. This technique is known as photon counting or spike counting~\cite{Akerib_2018_PRD} and is easy/feasible only at low $E$. More importantly than the slight improvement in energy resolution, at only the lowest energies ($<$10 keV), this reduces the leakage of background ER events into the WIMP (NR) region in S2~vs.~S1~\cite{Akerib_2016_Run03ReAnal}. All points plotted are defined as raw $\sigma/\mu$, but comparable results can be found with Gauss/skew centroids or medians.

Cyan lines represent S2, which, as seen before, can be better than S1 or even combined-$E$ scales, but only at $O$(1) keV. It is at least comparable, which is important given the historical use of only S1 for $E$ even in 2-phase TPCs with both channels, and continued usage for gas-less regions like the skin vetos of LZ and XENONnT. For S2, it is critical to understand the limitations, especially in low-mass-DM experiments like LBECA, where it is the only channel used. It can suffer from poor drift $e^-$ lifetime (impurities), incomplete extraction at a liquid-gas interface due to fields being too low, or both. The former effect (same as latter) is shown by the dash vs.~solid cyan. Even when lifetime and extraction are known, along with single-$e^-$ pulse size, low values lead to high S2 area variation, although the effect is muted above 50~keV. The 41.55~keV ``hiccups'' are $^{83m}$Kr, a combination of 2 decays, at 9.4 and 32.1~keV~\cite{Manalaysay_2010}. An inverse square root is not a good fit to the S2 or S1 alone and not included; it is due to incomplete accounting of quanta, also to linear noise flattening the curves. Such noise impacts higher $E$s more and is defined in detail in Section 4.3 of \cite{Dahl:2009nta} and Section III.A of \cite{szydagis2020investigating}. As seen in the $\sim$straight lines in log--log, a power law (often plus constant for noise) is reasonable for combined $E$ (blue) but below $\sim$5~keV even $E^{-\frac{1}{2}}$ breaks down, unless the power is free, preventing extrapolation from 10--100+~keV down to where behavior may even be non-analytic, and $E$ resolution is ill-defined.

The NEST combined scale is in solid blue in Figure~\ref{Fig2}a compared to LUX data from its first science run as black circles. LUX used the same combined scale, which again is clearly advantageous compared to single-channel methods, with $g_1=0.117 \pm 0.003$ and $g_2=12.1 \pm 0.9$~\cite{Akerib_2016_Run03ReAnal,Akerib_2016_CH3T}. The exception in the plot is the first point (2.8~keV) which should be compared to the optimized NEST in gold. The dashed and dotted blue lines are examples of further resolution improvements. First, by removing the linear noise, in MC, as doing so is certainly not as easy in data (noise proportional to the S1 and S2 areas representing, e.g., imperfect position corrections). This is modeled as only 1.4\% for LUX. It is typically $O$(1\%)~\cite{szydagis2020investigating,Dahl:2009nta}. Second, respectively, by simulating a uniform E-field, when the real field varied with position, though not significantly in LUX's first WIMP search run (180~V/cm average) and the (much larger) variation was taken into account in the second~\cite{Akerib_2017_Lucie}. The last few highest-$E$ points in data (black) do not overlap with MC due to not fully accounting for PMT saturation (S2 clipping) above 500~keV.

The advantages of the optimal scale (gold) disappear rapidly above 10~keV, comparing gold to blue. But a benefit is that at sufficiently high $E$ the rotating/re-weighting of a peak in S1 and S2 to find the optimal resolution results in a derivation of $g_1$ and $g_2$~\cite{Akerib_2014}. This abrogates the need for multiple peaks, arranged in S2~vs.~S1 (means) in what is known as the Doke plot. Such a plot is a straight line due to the anti-correlation between $N_{ph}$ and $N_{e-}$ for ER~\cite{Doke_2002} shown to work across at least four orders of magnitude in energy, and different fields~\cite{Akerib_2018_PRD,Dahl:2009nta,Dobi:2014wza,Aprile_2019_SignalAnal}. Alternatively, if studies of anti-correlation both within peaks and across peaks for a given analysis in a certain experiment are possible, then these two methods for deriving the S1 and S2 gains can serve as cross-checks, on top of NEST comparisons and known-spectrum reproduction such as that from tritium betas~\cite{Akerib_2016_CH3T}. As explained in the caption of Figure~\ref{Fig2}, the S1 weight $w_1$ is decreasing toward 0 with decreasing $E$s while $w_2$ increases to compensate; with increasing $E$s $w_1$ increases and $w_2$ decreases, while both $w$s asymptote to 1.0, as expected.

In the second plot plane (b), we focus on the mean instead of the width (resolution) demonstrating explicitly with both data from Compton scattering in black~\cite{Goetzke_2017} and our NEST MCs (despite significant uncertainties) that the thresholding effects raise the reconstructed energy significantly above the true value. This phenomenon becomes most prevalent in the sub-keV regime, however, where resolution becomes ill-defined due to individual photon and electron quanta becoming resolvable, generating a multiple-peak structure~\cite{luxcollaboration2016lowenergy}. The peaks become not just skewed-Gaussian, but entirely non-Gaussian, or even non-analytic~\cite{lenardo2019measurement}. For this reason, Figure~\ref{Fig2}a stops at 2~keV on the x-axis. Figure~\ref{Fig2}b continues below that, focused on mean (i.e., ratio of reconstructed over known energy) not width however. We switch to neriX from LUX here, as LUX does not have a relevant plot published with which we can compare, and did not have direct, quasi-monoenergetic measurements below 1~keV. (Nevertheless, due to similar $g_1$'s and $g_2$'s in these and most experiments the results should be quite general.) Data uncertainties are driven in x ($E$ axis) by finite resolution in the Ge detector used for independent energy determination, and in y by uncertainties in neriX's $g_{1}$ and $g_{2}$ (0.105 $\pm$ 0.003 phe/photon, 16.06 $^{+0.9}_{-1.0}$ phe/$e^-$). NEST uncertainties are large only at sub-keV, and are not statistical due to large simulations. Instead they are due to the uncertainty on how to define a central value, using a mean or median or attempted Gaussian fit, due to the multi-peak effect mentioned (photon and $e^-$~discretization).

At low $E$, benefits of not just a combined but optimally-combined (re-weighted) scale are great: not just a built-in erasure due to $w_2$ of the growing discrepancy between the reconstructed and real values of energy illustrated effectively in Figure~\ref{Fig2}b (see also neriX's Figure~7~\cite{Goetzke_2017}) but a reduction in width that was 50\% (relative) for $^{37}$Ar in both PIXeY and LUX. Lastly, as illustrated in Figure~\ref{Fig1}d, the shape becomes more symmetric at individual energies, with the skew nearly disappearing. While this matters more for monoenergetic ER peak searches (for axion-like particles or ALPs, bosonic WIMPs, etc.) a benefit for a WIMP search, or for any analysis in fact, is a better determination of the $g_1$ and $g_2$ through tighter windowing around single-energy calibration lines in 2D, in S2~vs.~S1, which can occur iteratively, reducing the errors on $g_1$ and $g_2$ (5--10\% typical) that often drive systematic uncertainties on both yield analyses and final physics results, especially in terms of S1 and S2 thresholds~\cite{Akerib_2016_CH3T,szydagis2020investigating,Akerib_2016_Run03ReAnal}. The only disadvantage is loss of the field-independence a combined scale usually has, as the yields change with field. As most experiments run at only one electric field however, that is not a true drawback.

\subsubsection{\textbf{High Energy: The MeV Scale (Neutrinoless Double-Beta Decay)}}

Far from the hard thresholds, we turn our attention next to 0$\nu\beta\beta$ decay. Searches for this require great resolution for good background discrimination, at $Q_{\beta\beta} = 2.458$~MeV for $^{136}$Xe specifically~\cite{Auger_2012_Xe136}. While resolution naturally improves with $E$ due to the greater numbers of quanta produced, effects such as PMT saturation and different noise sources, including position-dependent effects, become more prominent. While machine learning can help a great deal as done on EXO-200 especially with detector-specific idiosyncrasies~\cite{Delaquis_2018}, the analytic optimum scale becomes degenerate with combined $E$ above 0.1~MeV even already as illustrated earlier. Table~\ref{Tab2} reviews the resolutions achieved in actual experiments: projections of future performance, e.g., for LZ~\cite{Akerib_2020_0vBB}, are not included, in order to showcase only what has been demonstrated, or extrapolated with $\sigma/E \propto 1/\sqrt{E}$ (+~optional constant).

In EXO, in its references cited below, a richer formulation was adopted: $\sigma^2 = a + bE + cE^2$. It considers more detector noise sources. For $a = c = 0$, it simplifies to $\sigma^2 = bE$ ~or~ $\sigma/E = \sqrt{b/E}$.

\begin{table}[H]
\vspace{-5pt}
\caption{Survey of experimentally achieved energy resolutions at $Q_{\beta\beta}$ or very close to it (for instance, 2.6~MeV calibration). Entries marked with * are exceptions: extrapolations from power-law fits to much lower energies. Error bars are included whenever they were reported for the analyses. EXO made both hardware upgrades (Phase I, II) as well as many software enhancements.}
\centering
\tablesize{\small}
\begin{tabular}{ccc}
\toprule
\textbf{Experiment}	& \textbf{Resolution [\%] Gaussian 100*1$\sigma$/mean}	& \textbf{Uncertainty}\\
\midrule
XENON10~\cite{Aprile_2011_W14}	& 0.89*			& \\
XENON100~\cite{Aprile_2014_AnalXe100}			& 1.21*			& \\
XENON1T~\cite{Aprile_2020_resE}		& 0.80			& 0.02\\
EXO-200~\cite{Ackerman_2011} single-scatter only			& 4.5 for Q			& \\
EXO-200~\cite{Davis_2016} $ibid.$			& 1.9641			& 0.0039\\
EXO-200~\cite{Albert_2014_2vBB} & 1.84 (6.0-7.9 S1, 3.5 Q) & 0.03 \\
EXO-200~\cite{Auger_2012_Xe136}			& 1.67			& \\
EXO-200~\cite{Davis_2016,Albert_2017_Xe134,Auger_2012}			& 1.5820 or 1.6(0)			& 0.0044\\
EXO-200~\cite{2014_EXO-200_First,PhysRevC.92.015503,Albert_2016_Cosmo,Albert_2016_CPT,Leonard_2017}			& 1.53			& 0.06\\
EXO-200~\cite{Albert_2018_1p23,Anton_2019_EXO-200_Complete}			& 1.38 then 1.23 then 1.15			& 0.02 (last)\\
EXO-200~\cite{Delaquis_2018} & 0.94(1)-1.38(2) comb and 3.44(6)-4.08(4) for Q & \\
KamLAND-ZEN~\cite{zencollaboration2014results,Shirai_2018}		& 4.0-4.3	(GXe dissolved in liquid scintillator)		& \\
KamLAND-ZEN~\cite{Asakura_2016,Gando_2013,Gando_2012,Gando_2012_PRC}		& 4.2			& $\sim$0.2\\
KamLAND-ZEN~\cite{Gando_2016}		& 4.66			& \\
\bottomrule
\label{Tab2}
\end{tabular}
\vspace{-15pt}
\end{table}

The reader must be cautioned not to conclude that one technology (XENON is two-phase, but others single) is better, as fiducial mass and total exposure time, position resolution, overall background rate in the region of interest and self-shielding, and enrichment in $^{136}$Xe come into play. The XENON series of detectors have focused primarily on DM not 0$\nu\beta\beta$ and so were not enriched. Their intrinsically better resolutions are due not necessarily to a gas stage (converting Q into S2) but the use of PMTs with single-photon resolution, while EXO used silicon photo-multipliers (SiPMs) with poorer single-phe resolution (not needed at MeV energies) required for their lower radioactivity that is superior to even the custom PMTs for LZ/LUX and XENON~\cite{leonard2008systematic,Leonard_2017,Gallina_2019,Nakarmi_2020,Akerib_2013_PMT,Aprile_2015_PMT}. Lastly, we pass over GXe, for which there are many data points: NEXT has achieved better resolution than reported here, 0.1--0.3\% (0.30--0.74 FWHM) due to lower total-quanta Fano and lower recombination fluctuations~\cite{Alvarez:2012kua,Lorca_2014} both accounted for in NEST~\cite{szydagis_m_2018_4062516}. (The question of high mass vs.~superior resolution is beyond our scope.)

What we can do is perform detailed MC scans to predict the best potential LXe resolution, for 2458~keV, but differing conditions; real experiments measure it at a nearby $E$, e.g., 2615~keV from $^{208}$Tl. Validations are not overlaid, but we point to successes in predicting resolution for XENON1T~\cite{Aprile_2020_resE,Aprile_2020_Excess,szydagis2020investigating} and postdicting LUX. To narrow the enormous parameter space, infinite $e^-$ lifetime and 100\% extraction efficiency (or, all-LXe detector like n/EXO) are assumed, with 0\% noise in Q readout from the grids, but varying S1 noise level and wide E-field range for completeness. Second, an assumption is made of fixed medium $g_1 = 0.1$, a conservative baseline based on what is possible now, while NEST systematic uncertainty will be shown, stemming mainly from different assumptions for the Fano factor.

\begin{figure}[ht]
\includegraphics[width=0.99\textwidth,clip]{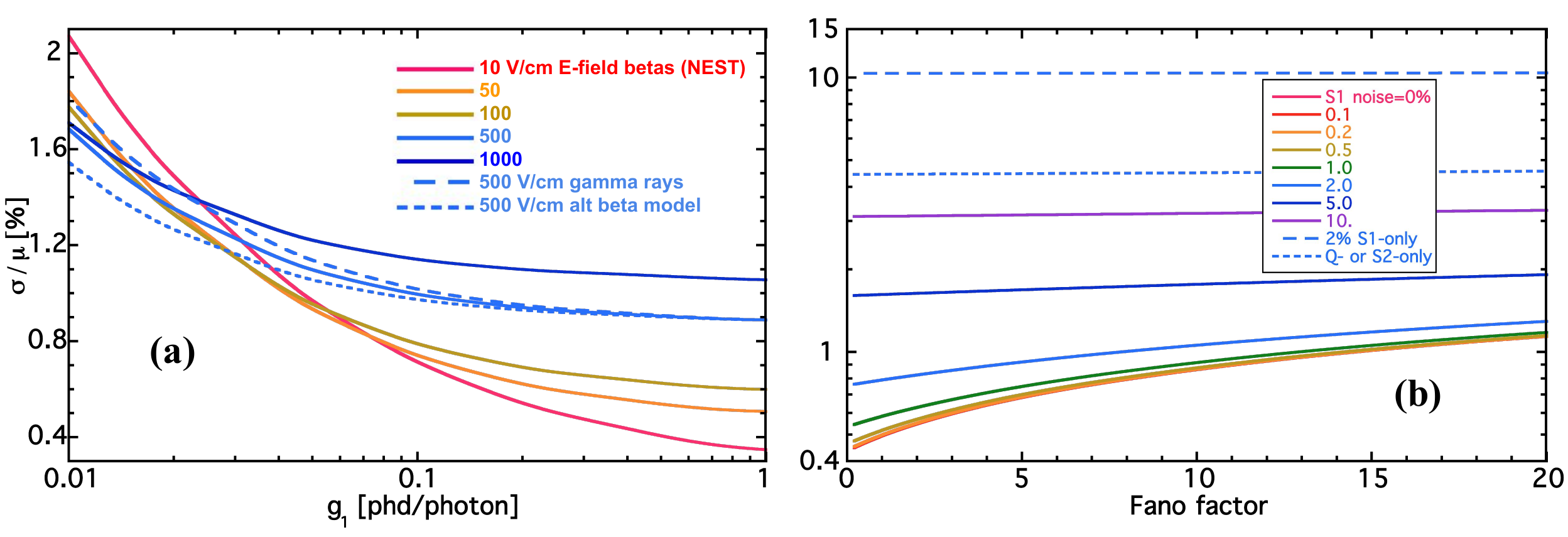}
\vspace{-0pt}
\caption{(\textbf{a}) 2.5~MeV resolution in LXe from NEST vs.~$g_1$ and fields. For one sample field, 500~V/cm, systematic uncertainty due to model choice shown: dashed cyan is the NEST electron recoil (ER) model based on photoabsorption gamma data, solid is the ER model from gamma Compton scattering and betas, and dotted line is a different beta model, based on \cite{Akerib_2020_GregRC14}. Differences are most significant at low $g_1$, where the exact $L_y$ value is more important. It is a key question for nEXO whether yields are more $\beta$- or $\gamma$-like. (\textbf{b}) The middle-of-the-road default beta model was selected, and $g_1$ and field frozen at 0.1 and 500~V/cm, and the resolution as a function of the Fano factor assumed is presented, for different levels of noise in the S1 (primary scintillation) signal. As on the left, combined $E$ used, except for the dashed line (S1) and dotted (S2 or Q) for comparison to more detectors, as close to the max (worst) possible.}
\vspace{-6pt}
\label{Fig3}
\end{figure}

Figure~\ref{Fig3}a is resolution dependent on $g_1$, from a pessimistic scenario of 1\%, all the way up to 100\%. Higher E-field is only better at low $g_1$, due to NEST's strictly empirical Fano factor $F_q$ increasing with field. It is unphysical, but needed to match data claimed to be de-noised or low in noise~\cite{APRILE1991177}. This is important, given the rush to achieve higher field for better resolution~\cite{Anton_2020} similar to the rush in the DM field, for lower leakage of ER backgrounds into the NR regime~\cite{PhysRevD.80.052010,araujo2020revised,luxcollaboration2020discrimination}. While $L_y$ and $Q_y$ are changing with E-field thus changing combined resolution, higher $g_1$ is naturally better, at least for non-zero field and combined $E$, due to more photons being collected. XENON1T, with its $g_1 \approx 0.13$ and field 120~V/cm, appears to have achieved close to the best possible for those values~\cite{Aprile_2020_resE} at 0.8\% (Table~\ref{Tab2}), also best overall. NEST's theory prediction of 0.7\% as best possible for XENON's $g$s and field rises slightly to match at 0.8 when applying XENON's $e^-$ lifetime and $e^-$ extraction efficiency~\cite{szydagis2020investigating}.

In Figure~\ref{Fig3}b is $E$ resolution's dependence on $F_q$, from a theoretical value~\cite{doi:10.1002/9783527610020.fmatter} (sub-Poissonian, $\ll$ 1) up to the largest experimental one, of Conti et al.~\cite{Conti_2003}. This governs standard deviation: $\sqrt{F_q N_q}$. While the best-fit (world data) NEST value, for 2.5~MeV and 500~V/cm, is 14 by default, we treat the Fano factor as free in Figure~\ref{Fig3}b, extending down to 0.2 due to NEST possibly absorbing detector-specific noises by mistake into the Fano value (even if this is not likely due to matching data across decades~\cite{APRILE1991177,Aprile_2020_resE}). All the various EXO-200 results can be explained, as being between $\sim$2 and 5\% noise levels in the detection of scintillation. The dotted cyan line holding steady at 4.4--4.5\% explains precisely the seminal result with Q-only resolution of 4.5\% in Table~\ref{Tab2}. It is not affected much by Fano factor, since when one considers only a single channel the recombination fluctuations, which move quanta between the scintillation and ionization signals, dominate~\cite{Angle_2008_1stXe10,Dahl:2009nta,luxcollaboration2020discrimination}. It is very similar at different fields, since in the minimally ionizing regime (as ER energy approaches and exceeds 1~MeV) $Q_y$ and $L_y$ asymptote to constants~\cite{Szydagis_2011,Akerib_2019_JonB,Akerib_2020_GregRC14}. The dashed line is too high to explain EXO's S1-only values, but S1 resolution improves with more light at lower E-field (higher E-field increases charge, at expense of light).

Regardless of whether it is achieved through ramping up $g_1$ (not unrealistic for the future given 100\% quantum efficiency QE devices~\cite{Superconducting} and high-quality reflectors~\cite{Kravitz2019MeasurementsOA}) or $F_q$ dropping to zero (it is not tuneable, at least not without doping of Xe with other materials; only feasible if the intrinsic value is already below Poisson, i.e., 1) the best possible value for resolution appears to be 0.4\%, a ``basement'' created by binomial fluctuations in excitation and ionization, combined with non-binomial recombination fluctuations~\cite{Dobi:2014wza,Akerib_2017_EvanP} if there is no noise (versus non-zero like 2\% example in pane b). Given realistic conditions, a more reasonable estimate of the min possible here is 0.6\% (close to GXe).

Even if the total number of quanta is higher than assumed here, due to $W$ being lower, as recently measured by EXO-200, $11.5 \pm 0.5$~eV~\cite{Anton_2020} as opposed to $13.7 \pm 0.2$~eV (Dahl~\cite{Dahl:2009nta}) or $13.4 \pm 0.4$~eV (Goetzke, neriX~\cite{Goetzke_2017}) or $13.8 \pm 0.9$~eV (Doke~\cite{Doke_2002}), then this basement is unlikely to change significantly, as even $F_q=0$ was explored above. This discrepancy, observed in the light not charge channel and thus unlikely to be due to charge amp calibration differences, may be due to SiPMs being more sensitive (relative to PMTs) to wavelengths other than VUV, such as infrared (IR scintillation has been observed in LAr~\cite{Escobar_2018}). While at 2nd order the fluctuation models in NEST would have to be revised, to 1st everything discussed here would remain the same, but with $g_1$ and $g_2$ estimates decreasing by $\sim$15\%.

\subsubsection{\textbf{Energy Reconstruction and Efficiencies for a Continuous Spectrum}}

In searching for either 0$\nu\beta\beta$ decay or the dark matter, a continuous-spectrum background can obscure any potential signal of beta decay or dark matter respectively, in addition to peaks in the background, or calibration peaks~\cite{Ara_jo_2012,PhysRevC.92.015503,Akerib_2015_BGs,Li_2017,Aprile_2018_BGs}. In our final ER analysis, we return to optimal combined $E$, but consider a non-monoenergetic spectrum. A new challenge appears, as the cross-contamination between bins in a histogram can make it difficult to separate the upward fluctuations of lower $E$s from simultaneous downward fluctuations from the higher bins.

This difficulty is exacerbated by the fact that energy resolution is not fixed, so this is not a flat or linear effect with which it is easy to deal analytically. The resolution of course degrades as energy goes to zero. As a result of the light and charge yields depending on energy, an additional problem is the fact a spectrum flat in (combined) energy is not flat in the S1 nor the S2, and not all background spectra are going to be flat. That being said, this is approximately true at low energies for DM searches in LXe TPCs, after the contributions from all background radioisotopes are summed together, from Compton plateaus, neutrinos, and/or beta spectra, as in \cite{Akerib_2020_GregRC14,luxcollaboration2020discrimination,Aprile_2020_Excess,szydagis2020investigating}.

A na\"{\i}ve optimization attempt for a uniform spectrum that allows both $w_1$, $w_2$ to vary distorts it more than normal. Better results are obtained fixing $w_1$, unlike before. LUX is the example again: it is NEST's default. A generic flat ER background is simulated from 0--20 keV in real energy (it should not be taken to represent the true backgrounds found within \cite{Akerib_2014,Akerib_2015_BGs}). An excellent analytic fit for the detection efficiency vs.~$E$ for a continuous spectrum is a modified Gompertz function suggested by a LUX collaborator: $10^{(2-m_1 e^{-m_2*E^{m_3}}-m_4 e^{-m_5*E^{-m_6}})}/100\%$, which addresses both low-energy threshold and high-energy cut-off. (Typical values: $m_1$ and $m_4$ are $O(10)$ while $m_2, m_3, m_6$ $O(1)$ and $m_5$ $O(10^4)$.)

\begin{figure}[th!]
\includegraphics[width=0.99\textwidth,clip]{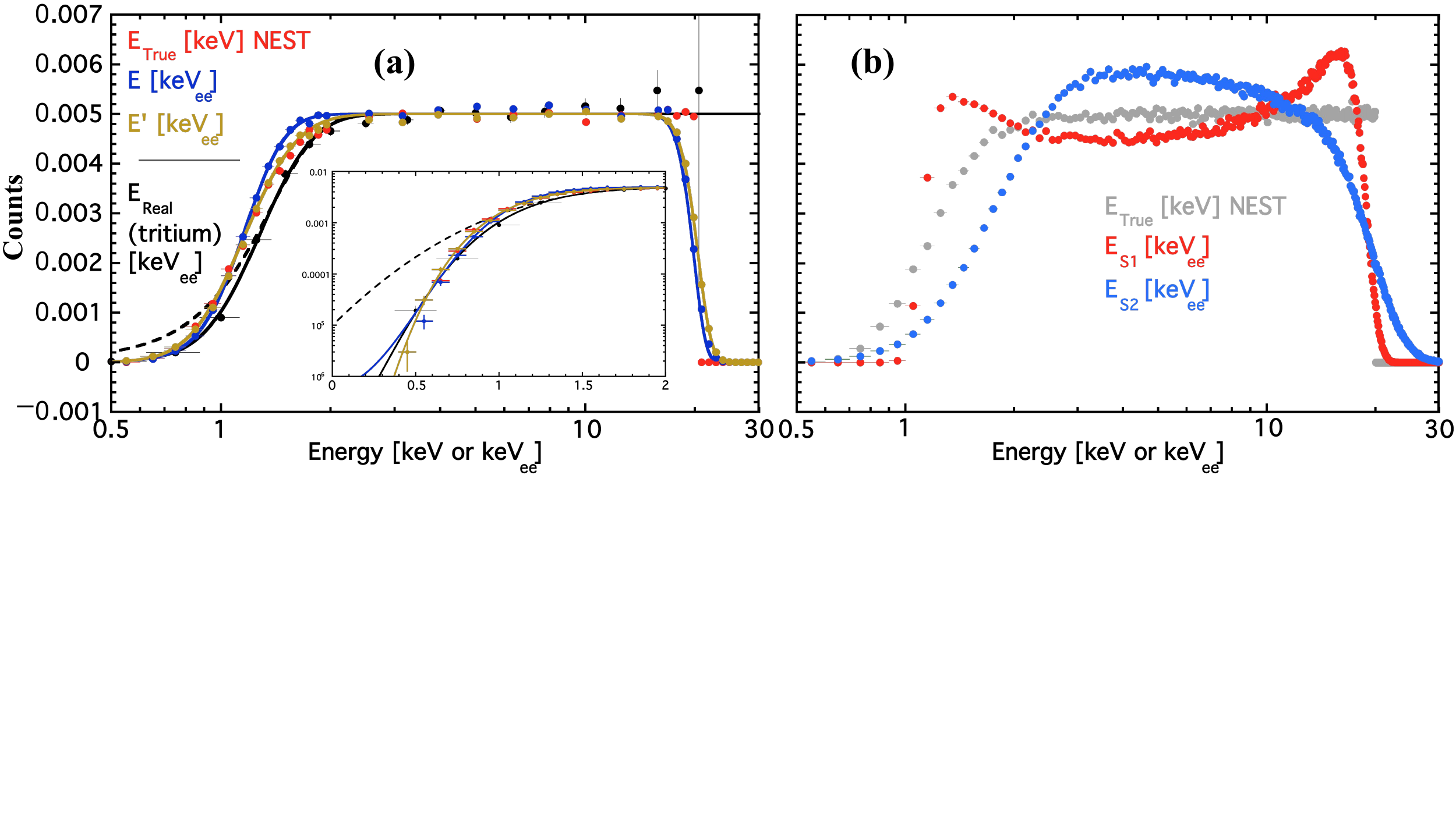}
\vspace{-107pt}
\caption{(\textbf{a}) Histogram of $10^6$ NEST uniform-$E$-spectrum events (0--20 keV true) binned using MC truth energies (red), reconstructed energy from the combined scale (blue), and re-weighted optimal reconstruction (gold). Gold outperforms blue even visually, correcting underestimation of efficiency sub-keV, plus overestimation near 1.5~keV (see the text for quantitative goodness of fit comparisons). Tritiated methane (CH$_3$T) is the black points, for validation against actual data. Its efficiency curve is markedly similar despite a non-flat spectrum. As it is continuous, the $E$ is still only reconstructed, not known to infinite precision as in MCs, although an attempt to empirically account for smearing was made by LUX~\cite{Akerib_2016_CH3T,Dobi:2014wza}. The solid black line is the Gompertz fit, superior to a more traditional $erf$, dashed, while the inset zooms on low $E$s for clarity, with a linear x-axis and log y now. (\textbf{b}) True $E$s now gray, but now compared to S1 (red) and S2-only (cyan) scales (Equations~(2) and (3)), with the former possessing unnatural peaks at left and right, and the latter grossly underestimating efficiency at keV scales. The default public $\beta$ model (NEST~v2.2.0) is used here but comparable results occur with $\gamma$s.}
\vspace{-5pt}
\label{Fig4}
\end{figure}

Figure~\ref{Fig4}a shows how the optimal scale, in gold again, is closer to the correct energies known from NEST MC in red, relative to traditional combined energy in dark blue again. $^3$H (tritium) betas are not flat in energy but their LUX trigger efficiency should be similar enough to a flat spectrum, so it is included in black to verify NEST's reasonableness~\cite{Akerib_2016_CH3T}. A $^3$H beta spectrum terminates ($Q_\beta$) at an 18.6~keV endpoint, but finite resolution causes the fluctuations around that energy. In gold, the $w_1$ is held constant at 1.0, but \mbox{$w_2 = 1.025 - e^{-\frac{E}{0.35}}$}, \textls[-10]{basically adjusting for the growing deviation between reconstructed and real energies as showcased in Figure~\ref{Fig2}b using an S-shaped curve asymptoting close to 1.0 (without a shrinking $w_1$, this weight $w_2$ has the opposite trend compared to Figure~\ref{Fig2}a).} $E$ in keV can stem from either the traditional combined scale, or MC truth, which in an actual experiment can be validated with a series of monoenergetic calibration peaks. The $\chi^2$/DOF = 2.65 for blue compared directly bin by bin with no fit to red at 0.5--17.5~keV, versus 1.68 for gold. Bin widths are 0.1~keV. The 50\% fall-off point at high $E$s (perfect step function in true $E$ in red) shifts from 19.5 to 20~keV, and is thus more accurate in gold compared to blue, forcing the endpoint smearing to be symmetric.

Figure~\ref{Fig4}b reiterates once more how distorted the S1/S2-only $E$ scales can be, in red/cyan. While Figure~\ref{Fig1} did show that S2 only can be the best for low-$E$ peaks, this is not the case for a continuum. Both S1/S2-only are non-uniform, despite the underlying spectrum being flat, and accounting for non-linearities in the underlying S1 and S2 yields, fitting quadratic not linear functions vs.~(true) $E$. The flat top should be 0.005 as in the truth spectrum, due to normalization: bin width over range = (0.1~keV)/(20--0~keV) = 0.005. S1 is area not spike, but phd~\cite{Faham_2015,Akerib_2018_PRD,Akerib_2017_Run04}. Despite this, there are unnatural peaks both low and high, caused by threshold and the max $E$ simulated (20~keV), respectively.

\subsection{Liquid Xenon Nuclear Recoil (Dark Matter Signal, and Boron-8 Background)}

Pivoting toward nuclear recoil, the first hurdle is that for this type of recoil the total number of quanta per unit energy is not fixed, unlike what was shown for ER (first by Doke et al.~at 1~MeV~\cite{Doke_2002}, and confirmed for energies of greater interest to DM experiments by Dahl~\cite{Dahl:2009nta}). This would seem to imply there is no anti-correlation between photons and electrons for NR and thus no benefit to using a combined energy scale for them. However, this does not appear to be the case, as the sum of quanta in actual data is well-fit by a power law, when combining all world data ever collected. This power law simply replaces the flat line (or general linear function but with no y-offset, if not dividing by energy: twice the energy means twice the $N_q$) that works so well for ER, given a fixed work function $W_q$ averaged over both flavors of quantum. This fit is related simply to the $L(E)$ from Equation~(1).

The mixing of units is possible, and depending upon whether one uses an S1-only, S2-only, or combined-energy scale, for NR the unit of keV$_{ee}$ can mean the beta, Compton, or photoabsorption event equivalent energy at which NR produces the same amount of S1, amount of S2, or the sum. In none of these three cases however is the conversion a simple constant or linear function, and can differ wildly, from keV$_{ee}$ being 2--10$\times$ smaller than keV$_{nr}$, depending also on energy~\cite{NESTWebSite}. The reason it is smaller is it takes less energy for ER to produce the same number of quanta compared to NR for the same energy deposit ($L < 1$). For the combined-energy scale, it is at least field-independent, as $L$ should not depend on field, only the recoil energy, and the $E$ resolution may be best via combination of information from both S1 and S2 again. (For additional clarity: some authors refer to $L$ as $f_n$~\cite{Sorensen_2011}.)

\begin{figure}[p!]
\includegraphics[width=0.85\textwidth,clip]{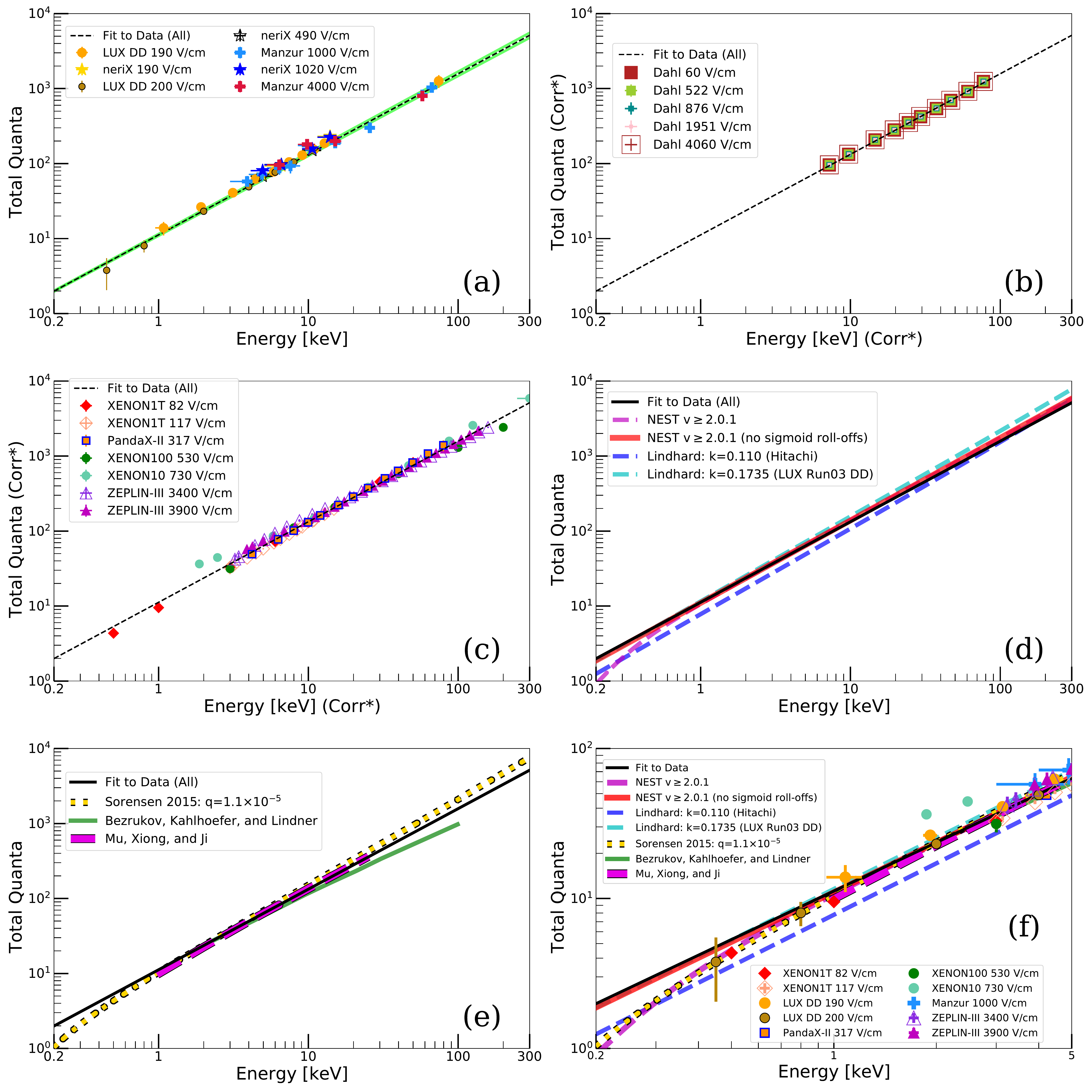}
\vspace{-0pt}
\caption{\small{All available world data on LXe NR at time of writing summarized in one place, with $N_{ph}$ and $N_{e-}$ combined into $N_q$. Where the $E$s did not match up (sometimes even within the same data set), a simple power law was used to spline (only interpolate not extrapolate) the numbers of photons first to add them to electrons. E-field does not cause measurable differences. (\textbf{a})~Only directly measured yields using angular measurements to determine $E$~\cite{Manzur:2009hp,luxcollaboration2016lowenergy,Aprile_2018_NeriXNR,Huang:2020ryt}. These are handled as more ``trustworthy'' in the community due to being quasi-monoenergetic analyses, and thus given the most weight in the fit even if there were fewer points. One simple power law appears to describe the data points across over three orders of magnitude in $E$, depicted as the black line, dashed or solid, within every plot pane. For (\textbf{a}), an uncertainty band was included (2$\sigma$ for clear visualization in green, not 1$\sigma$). (\textbf{b}) Dahl's thesis data from the Xed detector taken from broad-spectrum shape spline fits ($^{252}$Cf)~\cite{Dahl:2009nta}. *Corr(ected) on the y-axis refers to correcting the data in our global meta-re-analysis for effects often not known at the time of data-taking, such as the 2-photoelectron (phe) effect, or the extraction efficiency being much less than 100\% than the data-takers had originally estimated~\cite{Faham_2015,Edwards_2018,Xu_2019}. The former can lower the $L_y$ measurements, depending upon the analysis technique, while the latter raises $Q_y$ data points typically. The x-axis was ``corrected'' in the sense of energy estimates updated with a more modern combined-$E$ scale whenever possible. (\textbf{c}) More (indirect) measurements from continuous source bands, from XENON, ZEPLIN, and PandaX~\cite{Sorensen_2009,Sorensen_2010,sorensen2010lowering,Sorensen_2011,Horn_2011,araujo2020revised,Aprile_2013_LeffaQy,Aprile_2019_SignalAnal,Aprile_2019_Xe1TmcPLR,yan2021results}. Errors in data used whenever reported. (\textbf{d}, \textbf{e}) Review of models~\cite{szydagis_m_2018_4062516,Hitachi:2005ti,luxcollaboration2016lowenergy,Sorensen_2015,Bezrukov_2011,mu2013scintillation,mu2013ionization}. (\textbf{f}) Low-$E$ zoom of models, plus the data, as larger-sized points.}}
\vspace{-0pt}
\label{Fig5}
\end{figure}

Figure~\ref{Fig5} has all data available on $N_{ph}+N_{e-}$, from which we extract $L$. The plots suggest combined $E$ may still be beneficial even for NR, due to anti-correlation. The evidence is indirect, but strong: $>$300 data points from $>$20 experiments across nearly 2~decades were combined, respecting the systematics of each (typically driven by how well $g_1$ and $g_2$ were known). Remarkably, within uncertainty at least at the 2-sigma level the vast majority of the the hundreds of data points lie along the same straight line in log-log space. Publications reporting continuous lines stemming from, e.g., a modified NEST version or their own custom MCs are not ignored, but a few sample points at discrete $E$s are plotted. Figure~\ref{Fig5}'s plot style is similar to that pioneered by Sorensen and Dahl in~\cite{Sorensen_2011} as well as in later works.

Uncovering direct evidence of anti-correlation within NR is challenging: monoenergetic neutron (n) sources exist, but internal monoenergetic NR sources do not. The common sources used such as AmBe and $^{252}$Cf produce $O$(1-10)~MeV n's that lead to Xe recoils $O$(1--100)~keV for calibrating DM detectors. While neutrons can be background~\cite{Aprile_2013_nBG}, they are sub-dominant compared to ER~\cite{xenon100collaboration2013study,Akerib_2020_Gamma} and appear more commonly as the stand-in for DM used to calibrate WIMPs, being neutral particles that primarily scatter elastically, as DM should do~\cite{McCabe_2010,LEWIN199687}. The closest one can get to separable recoil energies comes from determination of neutron angle and double scattering as done in LUX using a D-D neutron generator, but even in this most optimum situation the direct testing of anti-correlation was not possible: $Q_y$, $L_y$ were reported at energies that did not match, and the former data included x-axis ($E$) error bars driven by uncertainty in angle, while the energy error for (single-scatter) $L_y$ data stemmed in turn from uncertainty from the $Q_y$ used to establish an {in~situ} S2-only $E$ scale~\cite{luxcollaboration2016lowenergy,Verbus:2016xhu}.

As this is not a paper presenting any novel models (in NEST for instance), we do not focus on breakdown of the total quanta into $N_{ph}$ or $L_y$ and $N_{e-}$ or $Q_y$, which numerous papers already discuss at great length~\cite{Wang_2017}. We also pass over the Migdal Effect, which could increase the light and/or charge at keV scales due to additional ER from initial NR; there is no evidence of its existence at present, but it is predicted to describe the behavior of electrons ``left over'' after a nucleus recoils~\cite{Aprile_2019_Migdal}. Additional phenomena like it are secondary when combining all individual channels. Returning our attention to Figure~\ref{Fig5}a--c, the empirical total number of quanta is described by the following power law (black line):

\vspace{-5pt}
\begin{equation}
N_q = \alpha E ^{\beta}, \mathrm{~where~} \alpha = 11.43 \pm 0.13 \mathrm{~and~} \beta = 1.068 \pm 0.003; \quad N_q/E = \alpha E ^{\beta-1} \sim 11 \frac{quanta}{keV}
\label{Eqn6}
\end{equation}

Not only can summed data of \cite{Dahl:2009nta,Aprile_2019_Xe1TmcPLR,luxcollaboration2016lowenergy,Aprile_2018_NeriXNR,Huang:2020ryt,Aprile_2013_LeffaQy,Sorensen_2009,Sorensen_2010,sorensen2010lowering,Angle_2011,Manzur:2009hp,Horn_2011,Hitachi:2005ti,Sorensen_2015,Bezrukov_2011,mu2013scintillation,mu2013ionization} be described simply with a power fit but with an exponent near 1, implying a nearly fixed number of quanta/keV, $\sim$11.5 ($cf.$~73 for ER, coming from $1/W$), i.e., fixed $L$ of $\sim$0.157. The conversion into $L$, re-arranging Equation~(1), is:

\vspace{-5pt}
\begin{equation}
L(E) = \frac{N_q}{E} W_q = \alpha E ^{\beta-1} W_q = (0.1566 \pm 0.0018) E^{0.068 \pm 0.003} \approx 0.1707 + 0.0012 E
\label{Eqn7}
\end{equation}

$W_q = 13.7 \times 10^{-3}$~keV is assumed here with no uncertainty as just an example, while the last term is the Taylor expansion to first order at 10~keV. In Figure~\ref{Fig5}d,e, the fit to $N_q$ from data is compared to several models, starting with the traditional Lindhard approach~\cite{Lindhard_1963,LEWIN199687}:

\vspace{-5pt}
\begin{equation}
L(\epsilon) = \frac{kg(\epsilon)}{1+kg(\epsilon)}, k = 0.133~\frac{Z^{2/3}}{A^{1/2}} \approx 0.166, g(\epsilon) = 3\epsilon^{0.15}+0.7\epsilon^{0.6}+\epsilon, \epsilon = 11.5~E~Z^{-7/3} \approx 0.00105 E
\label{Eqn8}
\end{equation}
\vspace{+1pt}

where Z = 54, A = 131.293 (average) for xenon. $\epsilon$ is known as the ``reduced energy.'' It allows dimensionless $L$ comparison across different elements. A Taylor expansion for Equation~(8) at 10~keV is $L(E) = 0.1790 + 0.0028 E$. At this $E$, the value of $L$ for the expansion of Equation~(7) is close, $<$5\% lower than the one for Lindhard (8). Equation~(7)'s linear approximation is lower for both of its terms, but this can be explained by bi-excitonic and/or Penning quenching, which increases with higher $dE/dx$, which occurs with increasing $E$s for keV NR. Colliding pairs of excitons may lead to de-excitation, and thus less S1~\cite{Hitachi:2005ti}. Some fraction of excitation may also be converted into ionization, adding to Q. Using the values in \cite{Manzur:2009hp} or \cite{Bezrukov_2011}, it is even possible to show that NEST, similar to the data fit, follows Lindhard closely above $O$(1~keV), if additional quenching is added, for photons (see \cite{NESTWebSite}).

This is remarkable as it was not expected that Lindhard would work even that high in $E$~\cite{Sarkis_2020,Hitachi:2005ti,Sorensen_2015}. Yet data, once summed, exhibit no significant deviation from the Lindhard model, down to sub-keV even. At higher $E$, the work of Hitachi~\cite{Hitachi:2005ti,PhysRevLett.97.081302}, who incorporates quenching, may be more appropriate; however, it can be approximated using Lindhard but with $k = 0.11$ (blue). Figure~\ref{Fig5}d also includes the fit to the LUX D-D n gun data alone (green) that agrees with standard Lindhard at the $1\sigma$ level between 0.7--74 keV, given $k=0.174 \pm 0.006$, after accounting for extra quenching, separately~\cite{Akerib_2018_PRD,luxcollaboration2016lowenergy}. Sorensen 2015 in yellow in Figure~\ref{Fig5}e assumes standard-$k$ Lindhard at high $E$s, and an atomic-physics motivated roll-off below 1~keV, with a free parameter $q$ (in same units as $\epsilon$) we chose to best match all contemporary data, $1.1 \times 10^{-5}$ or 10.5~eV. Only the min (blue) and max (green) $k$ (which is uncertain and can range from 0.1-0.2~\cite{Sorensen_2011,mu2013scintillation,Sarkis_2020}) easiest to justify are depicted (Figure~\ref{Fig5}d). $k= 0.166$ would be between, as would 0.14 from Lenardo et al., best fit to data as of 2014~\cite{Lenardo_2015}.

Any NEST version $\geq$2.0.1 has a power fit of $11 E ^{1.1}$ (red, Figure~\ref{Fig5}d) matching Equation~(6) with rounding, but more importantly different because of considering not just raw yields but $log_{10}(S2_c/S1_c)$ band means, and giving greater weight to the lowest-$E$ data, of greatest importance to a DM search~\cite{LEWIN199687} (previously $12.6 E^{1.05}$, before all statistical and systematic uncertainties were properly considered). The dashed purple line of Figure~\ref{Fig5}d is not just the power law but the full NEST that also since v2.0.1 to today includes sigmoidal corrections, separate in both $L_y$ and $Q_y$, to allow for modeling of NR violating strict macroscopic anti-correlation. These cause the $N_{ph}$ and $N_{e-}$ to realistically drop below the power law, conservatively accounting for non-Lindhard-like behavior below a few keV, and better matching data in this regime such as from the D-D calibration of LUX's second science run~\cite{Huang:2020ryt}. Nevertheless, all models agree very well at low $E$s, all of them extrapolating at 200~eV to 1--2 quanta (see Figure~\ref{Fig5}f). At $<$0.2~keV, NEST conservatively assumes 0 quanta, justifiable from first principles~\cite{Sarkis_2020,Sorensen_2015,NESTWebSite,szydagis_m_2018_4062516}.

For greater readability several models have been omitted from Figure~\ref{Fig5}d,e, which do not include the work of Sarkis, Aguilar-Arevalo, and D'Olivo~\cite{Sarkis_2020} nor of Wang and Mei~\cite{Wang_2017}. This is not to say their approaches are not valuable, but the former is markedly similar to Sorensen~\cite{Sorensen_2015} and to the complete NEST equation including a sub-keV roll-off despite starting with different formulae, while the latter is fit to the LUX data, and thus practically redundant with the dashed green. In Figure~\ref{Fig5}f we zoom in on energies $<$5 keV only, of greatest interest not only for lower-mass WIMPs, but any rest mass-energy WIMP due to the falling-exponential nature of NR from WIMPs, in the Standard Halo Model~\cite{McCabe_2010}. The cut-off in quanta may not be as sharp as expected due to the Migdal Effect adding more light at least~\cite{Akerib_2019_Migdal}. Below 3~keV $^8$B solar neutrino CE$\nu$NS is of great significance as well, interesting in its own right for the first detection of the recently measured coherent scattering~\cite{Akimov:2017ade,Akimov:2020pdx} but from solar neutrinos, and as a background to next-generation LXe-based WIMP experiments~\cite{Akerib_2020_LZ_ProjSens,thexenoncollaboration2020projected}.

In following the ER section, the next step should be discussion of resolution as a function of $E$. This has never been published for the combined scale with NR, however, to the best of our knowledge, except for plotting of the S1- and S2-only scales only once each as far as we know, by Plante~\cite{Plante:2012umc,Plante_2011} (but only at zero field, in a dissertation) and Verbus~\cite{luxcollaboration2016lowenergy,Verbus:2016xhu}, respectively. In both situations these were only quasi-monoenergetic reconstructions, tagging neutron scatter angle with a Ge detector in the former case, as typical for all $L_{eff}$ measurements, and \textit{in~situ} in the same Xe volume (LUX) in the latter. For NEST comparisons to both of these, see Figure~3 bottom in the LZ simulation paper~\cite{Akerib_2021_LZSim}. We also point to potential examples of threshold bias ``lifting'' $E$s above correct values (or, Eddington bias, for continuous spectra), as manifesting in light yields~\cite{doi:10.1002/9783527610020.fmatter,Bernabei:2001hk,Chepel_2006} seen much earlier for ER in Figure~\ref{Fig2}b but it is handled in later works~\cite{Verbus:2016sgw,Verbus:2016xhu} and thus not shown explicitly pre-correction.

Lacking truly monoenergetic peaks, continuous spectra present an opportunity still for contrasting the S1, S2, combined, and optimized-$E$ scales. Those from all known n sources are highly dependent on geometry, however; thus NEST is insufficient: a full-fledged Geant4 MC~\cite{AGOSTINELLI2003250,Allison:2006ve} would be required to model detectors. Instead, an example of a 50~GeV/$c^2$ mass WIMP will be shown with an unrealistically large cross-section of interaction (1 pb, and 1 kg-day exposure). While artificial, this illustrative example is valid given underlying assumptions for not just total quanta but individual photons and electrons, and resolution in both channels, verified by NEST comparison with data elsewhere~\cite{Szydagis_2011,Szydagis_2013,Lenardo_2015} and real experiments will be able to test the ideas presented in future NR calibrations, in XENONnT and LZ.

\begin{figure}[t]
\includegraphics[width=0.98\textwidth,clip]{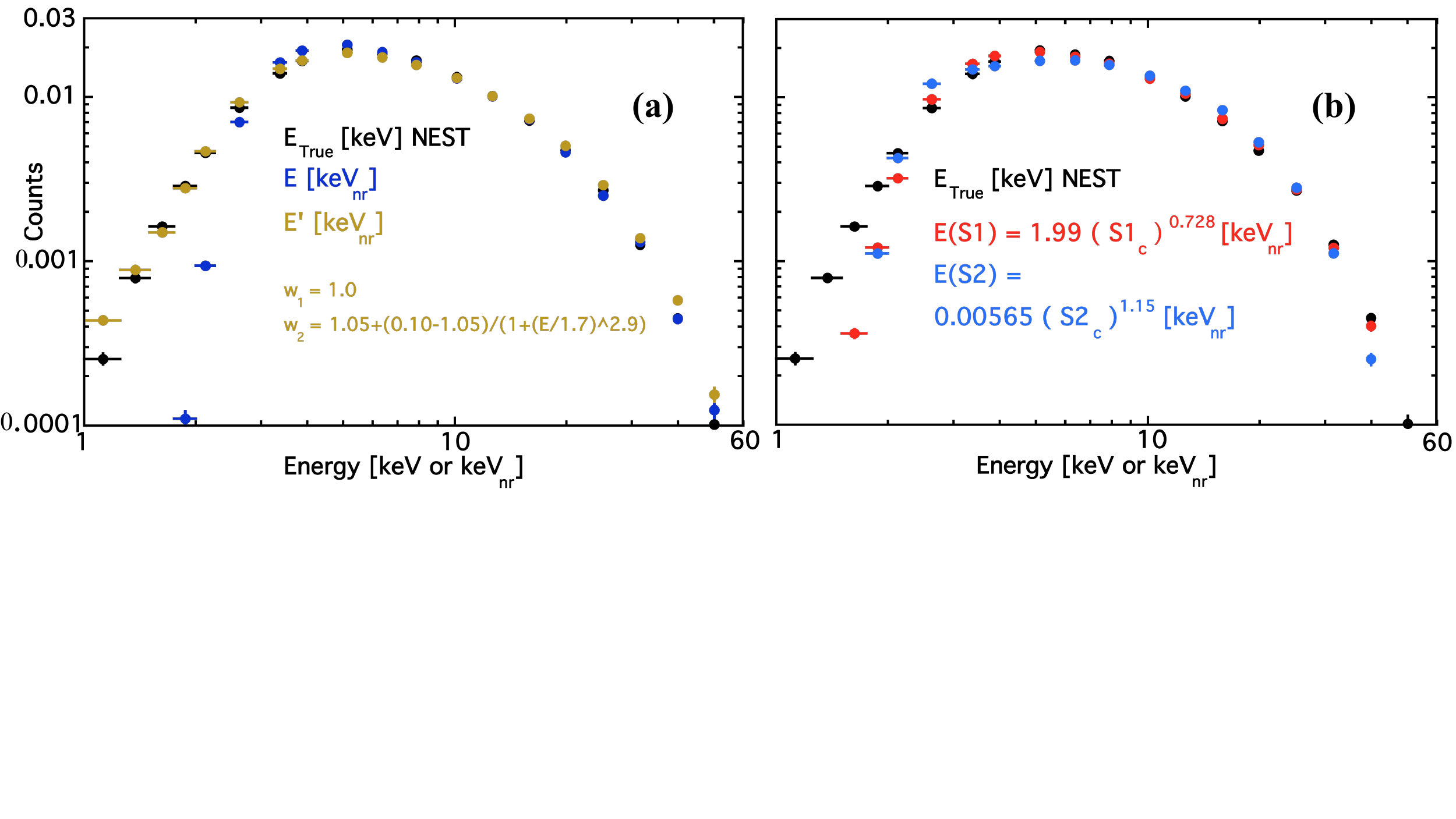}
\vspace{-105pt}
\caption{(\textbf{a}) Combined (blue) and optimal (yellow) scales for a 50 GeV standard weakly interacting massive particle (WIMP) spectrum, the MC truth for which is black. Above 2 keV, bins are omitted in log fashion for clarity. LUX detector parameters (Run03) used as earlier for similar ER plot, again for illustrative purposes ($\sim$180~V/cm). Yellow corrects for sub-keV underestimation, and an overestimation at several keV, comparing yellow to black and blue to black (these effects can change from detector to detector, with signal shape). While a distinct functional form vis-\`a-vis ER, $w_2$ is again S-shaped. Better results may be possible with lower $w_1$, kept fixed at 1.0 again here for simplicity. In actual data, with ER and NR mixed, backgrounds with potential signals, it is impossible to know {a~priori} if applying keV$_{nr}$ or $ee$ is more appropriate by event. (\textbf{b})~True spectrum is repeated, but now compared with S1/S2 in red/cyan. Power laws in legend are crude, detector-specific approximations, but can be converted with $S1_c=g_1 N_{ph}$, $S2_c=g_2 N_{e-}$ ($g_1=0.117$, $g_2=12.1$). For why both powers are $\sim$1, see Equation~(6).}
\vspace{-0pt}
\label{Fig6}
\end{figure}

Figure~\ref{Fig6} is a repeat of Figure~\ref{Fig4}, but now it's for NR. The WIMP spectrum is more relevant than a uniform-$E$ spectrum would be, as even for a massive (multi-TeV mass-scale) WIMP flat is a poor approximation to what is inherently an exponential spectrum, falling as energy increases. The drop off on the left is of course caused as before by a combination of separate threshold effects that remove the lowest-energy S1s and S2s. The optimal scale shows improvement again over combined, but by itself combined $E$ is not dissimilar from S1- or S2-only for NR, because the lower-quantum/area signals are dominated by detector specifics such as (binomial or Poisson) light collection efficiency. S1 only is most common, used in every experiment starting with the seminal XENON10 result~\cite{Angle_2008_1stXe10} due not to a better energy reconstruction, but signal (NR) vs.~background (ER) discrimination~\cite{Dahl:2009nta,luxcollaboration2020discrimination}. S2 only may be as good for discrimination if not better though, according to Arisaka, Ghag, Beltrame et al.~\cite{ARISAKA201251}.

\subsection{Liquid Xenon Summaries}

The key points of the LXe (ER) section (with detector-specific caveats) are:
\vspace{-5pt}

\begin{itemize}[leftmargin=*,labelsep=5.8mm]
\item A combined scale reconstructs monoenergetic ER peaks best for DM/$\nu$ projects, but below 3~keV at least this is not true according to an $^{37}$Ar study with S2 only best (outperforming S1 as well) if $e^-$ lifetime is high. A combination can be established with two numbers, S1 and S2 gains, leading to a 1D histogram (XENON/LUX style) or equivalently 2D rotation angle (Conti/EXO method).
\item An optimal weighting of S1 and S2 can result in better resolution than simple combined energy, down to $O(1)$~keV even, and mitigation of threshold bias and skew. Higher, the best resolution occurs when the weights applied to the S1 and S2 are $1/g_1$ and $1/g_2$, but machine learning is likely to outperform analytic methods, if more parameters (beyond S1, S2) are considered.
\item For neutrinoless double-beta decay, $O$(1\%) resolution has been achieved in the relevant energy range by a multitude of different experiments and technologies, while the best feasible may be 0.4--0.6\%, in liquid, which may be limited by a Fano factor (often confused with recombination fluctuations) that is higher than in gas.  No one experiment has yet reached its full potential.
\item For a continuous ER spectrum, the combined scale is a clear winner over S1-only and S2-only alike, at least for a uniform energy distribution (uniform in neither S1 nor S2, as $L_y$ and $Q_y$ are functions of energy, not flat). Optimization with re-weighting is still possible, just in a different manner than done for monoenergetic peaks, because of cross-contamination between bins.
\end{itemize}

\vspace{-5pt}
Next we summarize NR; there is good agreement on total yield from different experimentalists.
\vspace{-5pt}

\begin{itemize}[leftmargin=*,labelsep=5.8mm]
\item While impossible to obtain from truly monoenergetic lines, the summation of separate $N_{ph}$ and $N_{e-}$ data sets results in strong evidence of NR anti-correlation akin to ER's and no statistically significant difference from Lindhard even sub-keV, at least given additional high-$E$ quenching.
\item Despite the point above, the advantages of a combined scale are not significant compared to the S1-only default (but S2 comparable) as so much $E$ is lost to heat ($>$80\%) decreasing pulse areas.
\item An optimized combination scale, which corrects for order-of-magnitude discrepancies in efficiency below 1~keV, is still best, but likely requires fine-tuning by energy spectrum. It is also likely to be highly detector-dependent and only important after a WIMP discovery is made, to fit the mass and cross-section the most precisely. A uniform spectrum is a bad approximation in any case.
\end{itemize}

\subsection{Liquid Argon Electron Recoil}
\vspace{-15pt}
\subsubsection{\textbf{Low Energy: keV-scale (Dark Matter Backgrounds/Calibrations) Monoenergetic~Peaks}}

For ER in LAr, the best example of a low-energy calibration line is the $^{83m}$Kr peak, at 41.5~keV, commonly used to calibrate both LXe and LAr experiments, but in this latter case there exists no evidence of the yields depending on the separation time between the individual 32.1 and 9.4~keV peaks~\cite{Lippincott_2010}, unlike in LXe~\cite{Manalaysay_2010}, so the MC comparison is easier. The 2.82~keV electron captured from $^{37}$Ar has also been studied in LAr but most commonly at zero electric field in single-phase (liquid) detectors, making it a less ideal choice for complete NEST comparisons to S1, S2, and combined-$E$ histograms~\cite{Kimura_2020_PRD}. For a WIMP search, ER is again the main background, due to $^{39}$Ar in DarkSide-50 (DS-50) at Gran Sasso~\cite{Agnes_2015}, DEAP-3600 (and formerly CLEAN precursors) at SNOLab~\cite{3600collaboration2018results,Gastler_2012}, and ArDM (at Canfranc). Underground Ar depleted in this isotope reduces the background, but it remains dominant~\cite{Galbiati_2008}. In neutrino experiments, ER is the signal, via neutral-current, charged-current, and elastic-scattering interactions. In this section, we begin by focusing on dark matter at the keV scale, later moving on to cover the GeV scale more relevant for accelerator neutrino experiments like DUNE~\cite{Abi:2020loh}, with the intermediate scale at MeV also important for supernova or solar neutrinos~\cite{microboonecollaboration2020continuous,Abi:2020lpk}.

\begin{figure}[hp]
\includegraphics[width=1.7\textwidth,clip]{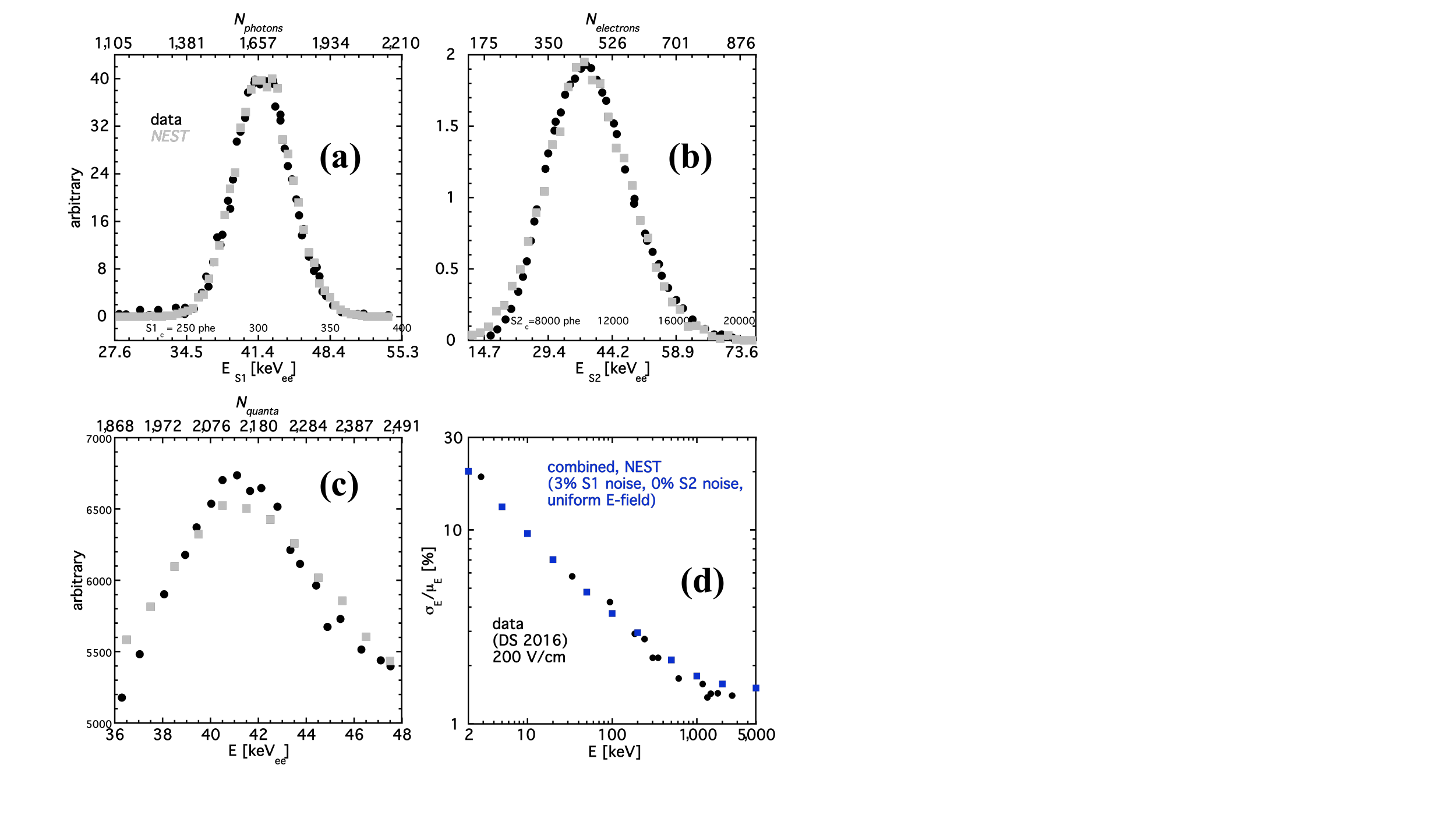}
\vspace{-40pt}
\caption{Reconstruction of the 41.5~keV $^{83m}$Kr peak in DarkSide (DS)~\cite{Pagani:2017} compared to NEST, with data as black dots, NEST gray squares. For clarity, no fits overlaid, although most data are $\sim$Gaussian. (\textbf{a})~S1-only $E$ scale most common for LAr-based DM detectors~\cite{Agnes:2016mgq}. Our work/conclusions for such a scale should apply not only to time projection chambers (TPCs) where ionization $e^-$s are drifted but also to 0~V/cm 1-phase detectors, where these electrons fail to recombine to add to the S1 in LAr for both ER and NR~\cite{Kimura_2019,Kimura_2020_PRD,Kimura_2020} just as in LXe~\cite{Szydagis_2011}, with the $L_y$ having the same shape vs.~$E$ as for non-zero fields as field goes to 0 in TPCs (see Doke et al.~as well as Wang and Mei for possible reasons~\cite{DOKE1976353,Doke_1988,Doke_2002,Wang_2017}). (\textbf{b})~S2-only, with the slight right-hand asymmetry nearly reproduced by NEST~\cite{Hackett:2017jnd}. (\textbf{c}) Combined-$E$ scale, now standard at least in DS. Optimization as done in LXe could be possible, but not shown. Resolution is poorer than in (\textbf{a}) instead of better due to poor S2 resolution in (\textbf{b}) and that S2 is being combined with S1, on top of $^{39}$Ar. This is explained in the text, as likely due to lack of XY correction creating noise not correlated with S1. For $E$s below 1 MeV, like here, this is not likely due to delta rays not being simulated by NEST by itself, as no S2 noise (and little in S1) is needed in the next plot (\textbf{d}) and S1 is not as wide as S2 in (\textbf{a}), while delta rays would affect both. Combined $E$, canceling them, should still be better in general in Ar as in Xe. (\textbf{d}) Resolution vs.~$E$ for monoenergetic calibrations/backgrounds studied on DS~\cite{Pagani:2017}. Only one set of resolutions covering a broad $E$ range was found (black) to compare to combined $E$ from NEST (blue).}
\label{Fig7}
\vspace{-0pt}
\end{figure}

Predictions of NEST's LAr ER model in comparison to experimental data are shown in Figure~\ref{Fig7}. The source of data here is DarkSide (DS), specifically several PhD theses of its students~\cite{Hackett:2017jnd, Agnes:2016mgq, Pagani:2017}. For this experiment, $g_1 = 0.18 \pm 0.01~$phe/photon~\cite{Hackett:2017jnd}, higher than within LXe detectors likely because of conversion of VUV photons into visible light using wavelength shifter followed by detection in high-QE visible-light SiPMs/MPPCs/APDs (Silicon Photo-Multipliers or Multi-Pixel Photon Counters or Avalanche Photo-Diodes). Another source gives this as $g_1 = 0.1856 \pm 0.0007_{stat} \pm 0.0008_{syst}$~\cite{Pagani:2017}. On the S2 side, $g_2 = 27 \pm 9$~phe/electron from \cite{Hackett:2017jnd} or $29.2 \pm 0.5_{stat} \pm 1.6_{syst}$ according to \cite{Pagani:2017}. Note that the gain factors $g_1$ and $g_2$ are called $\epsilon_1$ and $\epsilon_2$ by DS; others call them $\alpha_1$ and $\alpha_2$.

To most closely center NEST with respect to DS data in Figure~\ref{Fig7}, we set $g_1 = 0.181$ and $g_2 = 23.8$, both well within the above uncertainties. Only one example electric field was studied, of 200~V/cm, comparable to the electric fields studied earlier for LXe. The work function or $W_q$ was assumed to be 19.3~eV, a value justified later when discussing reconstruction for neutrino detectors. Reconstructed energy is again keV$_{ee}$ as opposed to keV (without the subscript) for known individual energies.

Unlike in LXe, the S1-only energy scale does not necessarily perform the poorest. Our particular $^{83m}$Kr example has an energy resolution $\sigma/\mu$ of 6.5\%, in Figure~\ref{Fig7}a. The S1-based scale is more reliable in LAr due to its greater linearity, wherein the light yield for different betas and gamma rays is quite flat in energy, starting at 40-50~photons/keV and falling as the electric field rises and more energy goes into charge production~\cite{Alexander_2013, Sangiorgio_2013, Aprile_2016, Zani_2014, Lippincott_2010, Agnes_2018, Doke_2002, Doke_1988}. Next, in Figure~\ref{Fig7}b, the S2-only scale is plotted, leading to a resolution of 25\%. Reproducing this large width was done artificially by setting the linear noise in the S2 channel to 25.0\% (only 2.9\% for S1 to match that width, an effectively negligible value). Compare this to 6.0\% (S2, so high due to $e^-$trains/tails/bursts~\cite{akerib2020investigation}) and 1.4\% (S1) assumed by default for LUX Run03 within NEST. We interpret the cause of the large value for S2 in LAr as stemming from a lack of full 3D corrections in the initial DS analysis. There is also a clear offset in skew compared to the MC which may come from DS-specific effects difficult to fully capture in a custom MC like NEST that does not account for all detector idiosyncrasies. No (statistical) errors are depicted as they are negligible from this high-statistics calibration.

Figure~\ref{Fig7}c shows a combined resolution of 7.5\%, not improving on the S1-only resolution, but still a marked improvement over S2-only, suggestive of anti-correlation of charge and light in LAr (more robustly established later in this section). The agreement is even worse with MC here, however, than in the S2 plot, but this can be explained by the peak being from a different analysis from (a) and (b), with the Kr calibration sitting on top of a large background, from $^{39}$Ar inside natural Ar, which we did not model (this is the only example of a DS combined-energy peak we were able to locate). In the original source the peak was not centered on 41.5~keV$_{ee}$, likely due to a systematic offset in $g_1$ and/or $g_2$ (see contradictory values above) but we had no difficulty in NEST with this. For ease of comparison for at least the width, 2.5~keV$_{ee}$ was added to data, to force alignment to a \mbox{41.5 keV(ee)} mean~\cite{Pagani:2017}.

Figure~\ref{Fig7}d is the combined-$E$ resolution vs.~$E$. NEST is blue. The data have full position corrections from high-statistics $^{83m}$Kr calibrations~\cite{Hackett:2017jnd,Agnes:2016mgq,Pagani:2017} as in LXe. They are most crucial for S2, in XY (or radius and angle) as Z is already handled by electron lifetime. NEST points for comparison are at semi-log steps in $E=$ 2--5000~keV. While retaining the 3\% S1 noise term from earlier, they have had the 25\% S2 noise from (b,c) removed, so they can be effectively treated as close to the MC-predicted min possible, at least for the given (DS-50) combination of $g_1$, $g_2$, and E-field. The combined-$E$ resolution drops from the 7.5\% in (c) to below 5\% in this case at 41.5~keV in (d). The empirical S2 noise term was thus likely accounting for imperfect correction for the (2D) position-dependent S2 light collection.

There is no equivalent of Figure~\ref{Fig2}b for LAr, as no example of Eddington-like bias could be found. An opposite effect (lower instead of higher reconstructed $E$) is recorded in Figure~3.11 of \cite{Pagani:2017}. However, this can easily be interpreted as $E$ ``loss'' into charge when using only S1, which while more linear in LAr compared to LXe, is still not a fixed flat $L_y$ at all energies, not even at null electric field~\cite{Kimura_2020_PRD,Agnes_2018_PRD}.

\subsubsection{\textbf{High Energies: The MeV and GeV Scales (Neutrino Physics)}}

In moving from the keV scale of DM experiments to the GeV scale of the accelerator neutrino experiments, we study the MeV scale as a boundary case. A 1~MeV beta was chosen as an example of an ER interaction energy just beyond what is considered relevant for a DM search, but on the other hand, at the extremely low-energy end for neutrino physics, potentially just barely above threshold in an experiment like DUNE~\cite{Weinstein_2020}. This is of the same order of magnitude but just slightly above the $^{39}$Ar beta spectrum endpoint (565~keV), and is also near the same energies (976~keV and 1.05~MeV internal-conversion electron and gamma ray, respectively, from $^{207}$Bi) studied by Doke et al.~in their seminal 1989--2002 papers~\cite{Doke_2002, Doke_1988}. Our own re-analysis shows that if one sums $L_y$ and $Q_y$ vs.~electric field, it is a constant number of quanta/keV within experimental uncertainties, and consistent with $1~/~W_q$, given reasonable assumptions on $g_1$ and $g_2$.

\begin{table}[h!]
\vspace{-5pt}
\caption{Best (lowest) possible resolution of a 1 MeV electron recoil at 0.5~kV/cm as a function of the S1 photon detection efficiency $g_1$ and wire noise (in semi-log steps). Each entry is averaged over 10$^4$ simulations in stand-alone NEST, with the effect of delta rays approximated analytically (based on G4). All photons and electrons are included from the interaction, which is thus being treated as a single site.}
\centering
\tablesize{\scriptsize}
\begin{tabular}{ccccccccc}
\toprule
\textbf{$g_1 [\%]$} & \textbf{0\% $Q$ noise} & \textbf{1\%} & \textbf{2\%} & \textbf{5\%} & \textbf{10\%} & \textbf{20\%} & \textbf{50\%} & \textbf{100\%} \\
\midrule
0.001 & 0.46 & 1.10 & 2.05 & 4.99 & 10.0 & 20.1 & 46.0 & 62.5 \\
0.002 & 0.46 & 1.10 & 2.05 & 4.99 & 10.0 & 20.1 & 46.0 & 62.5 \\
0.005 & 0.46 & 1.10 & 2.05 & 4.99 & 10.0 & 20.1 & 46.0 & 54.6 \\
0.01  & 0.46 & 1.10 & 2.05 & 4.99 & 10.0 & 20.1 & 40.5 & 46.6 \\
0.02  & 0.46 & 1.10 & 2.05 & 4.99 & 10.0 & 20.1 & 34.1 & 42.4 \\
0.05  & 0.46 & 1.10 & 2.05 & 4.99 & 10.0 & 17.8 & 28.9 & 31.5 \\
0.1   & 0.46 & 1.10 & 2.05 & 4.99 & 10.0 & 14.5 & 22.1 & 22.1 \\
0.2   & 0.46 & 1.10 & 2.05 & 4.99 & 8.78 & 12.7 & 15.6 & 15.6 \\
0.5   & 0.46 & 1.10 & 2.05 & 4.99 & 7.12 & 10.0 & 10.0 & 10.0 \\
1     & 0.46 & 1.10 & 2.05 & 4.21 & 6.34 & 7.08 & 7.08 & 7.08 \\
2     & 0.46 & 1.10 & 2.05 & 3.55 & 4.99 & 4.99 & 4.99 & 4.99 \\
5     & 0.46 & 1.10 & 1.79 & 3.08 & 3.14 & 3.14 & 3.14 & 3.14 \\
10    & 0.46 & 1.10 & 1.47 & 2.20 & 2.20 & 2.20 & 2.20 & 2.20 \\
20    & 0.46 & 0.87 & 1.28 & 1.53 & 1.53 & 1.53 & 1.53 & 1.53 \\
50    & 0.39 & 0.67 & 0.91 & 0.91 & 0.91 & 0.91 & 0.91 & 0.91 \\
100   & 0.21 & 0.57 & 0.57 & 0.57 & 0.57 & 0.57 & 0.57 & 0.57 \\
\bottomrule
\label{Tab3}
\end{tabular}
\vspace{-5pt}
\end{table}

An energy of 1~MeV is also associated with a track length at the border of position resolution in an experiment like DUNE, $<$1~cm on average, creating ``blips'' instead of obvious tracks, straddling the point-like interactions observed in DM detectors vs.~the track-like nature of most interactions in LAr TPC (LArTPC for short) neutrino experiments. Its low $dE/dx$ is also at the cusp of the minimally ionizing regime, making a 1~MeV electron a Minimally Ionizing Particle (MIP), like electrons from GeV-scale neutrino interactions prior to showering~\cite{Aguilar_Arevalo_2010}. In neutrino detectors there is no $g_2$ however, as electron charge is directly measured, as on EXO, by wire planes instead of S2. They are one-phase liquid TPCs with unit gain for charge readout. Table~\ref{Tab3} scans different levels of $E$ resolution associated with noise from the wires, showing best (lowest) resolution for different values of $g_1$. Combined-$E$ and even S1-only resolution appear in the table, for high $g_1$ paired with high wire noise.

Resolution in a DUNE-like detector can be halved already at $g_1=0.02$, i.e., 2\%, for 10\% wire noise. Such noise, starting even at 1\%, comprises Q-only resolution almost entirely. For 0\% wire noise, the 0.46\% resolution is driven by excitation and recombination, both contributing binomial fluctuations (unlike in LXe). $F_q$ drives combined-$E$ results. Its theoretical value 0.1 was assumed, given no other data~\cite{DOKE1976353}. For S1 only, a $g_1$-based binomial drives what is possible. 0\% noise was assumed for S1.

Unfortunately, large LArTPC neutrino experiments like DUNE will achieve much lower values of $g_1$~\cite{Acciarri:2016crz}, though $g_1=2\%$ is already much lower than any Xe or Ar DM experiment has achieved. At higher levels of noise, still realistic for future experiments but energy-dependent, even lower $g_1$ suffices for making combined energy superior to Q-only. When reading this table from the top down, if the value starts changing this means that the minimum resolution being quoted is from the combination of scintillation and ionization, and no longer just the ionization. If reading across: when the values stop changing that means (at high $g_1$) an S1-only scale is best, as at higher levels of wire noise not only does the Q-only energy scale become unreliable, but the benefit in utilizing the anti-correlation of charge and light washes out for the combined scale, leaving the S1-only scale.

The small-scale LArTPC R\&D detector LArIAT has investigated the claim the combined-$E$ scale, making use of both charge and light, is more precise~\cite{Acciarri:2019wgd,Foreman:2019dzm}. While not targeting the removal of the wire noise, it does effectively cancel out the exciton-ion and recombination fluctuations, shown using a sample of Michel electrons (from muon decay) at a scale of tens of MeV. Here, the definition of combination is more appropriately set for neutrino interactions using differential energy loss along the particle track, updating our earlier equations, starting with (1) but with $L=1$ assumed as for all ER:

\vspace{-15pt}
\begin{equation}
E=(L_y E+Q_y E)W_q = (\frac{dS}{dE}E + \frac{dQ}{dE}E)W_q \rightarrow \frac{dE}{dx} = (\frac{dS}{dE}\frac{dE}{dx}+\frac{dQ}{dE}\frac{dE}{dx})W_q = (\frac{dS}{dx}+\frac{dQ}{dx})W_q,
\label{Eqn9}
\end{equation}

where the number of photons $N_{ph}$ and number of electrons $N_{e-}$ are replaced for the first step by $L_y$ and $Q_y$, specific yields per unit energy, each multiplied by energy.

Next, to align our terminology with what is more common in the neutrino field we rewrite $L_y$ as $dS/dE$ and $Q_y$ as $dQ/dE$~\cite{Adams_2020} (instead of $N_{ph}/E$ and $N_{e-}/E$, with $S$ and $Q$ standing respectively for scintillation and charge, the quantum values). $S$ is used for scintillation instead of $L$ to avoid confusion with Lindhard. The formula above is still not what is most commonly used in the field. Instead, that is:

\vspace{-5pt}
\begin{equation}
\frac{dE}{dx} = \frac{dQ}{dx}\frac{dE}{dQ} = \frac{dQ}{dx} Q_y^{-1}(E,\mathcal{E}).
\label{Eqn10}
\end{equation}

This is equivalent to the S2-only scale used in dual-phase TPCs from Equation~(3) where $E = N_{e-}/Q_y(E,\mathcal{E})$, except divided by $dx$ and with $g_2$ = 1, given no need for extraction from liquid to gas, nor any photons created by electrons as a secondary process in gas (S2, i.e., electroluminescence). Although, the electron lifetime must still be high, ideally much larger than the full drift time across a TPC, and also well-measured, so that the Q can be corrected in the same way as S2. $Q_y$ (or $N_{e-}/E$) is often parameterized with Birks' Law~\cite{BIRKS1964269,Szydagis_2011,Agnes_2018_PRD} in terms of $dE/dx$ instead of $E$:

\vspace{-5pt}
\begin{equation}
Q_y(\frac{dE}{dx},\mathcal{E}) = A \frac{1/W_i}{1+(k_B/\mathcal{E})(dE/dx)} = A \frac{Q_0(E)/E}{1+(k_B/\mathcal{E})(dE/dx)},
\label{Eqn11}
\end{equation}
\vspace{-10pt}

where $W_i$ (sometimes called $W_e$) is not the same as the work function $W_q$ defined much earlier, but trivially related. Being defined as $E/N_i$, not $E/(N_{ex}+N_i)$, the convention of ionization $W$ is related to the overall or total work function by $W_i = (1+N_{ex}/N_{i})W_q$. ($N_{ex}$/$N_i$ are excitons/ions).

\textls[-20]{For LAr, this means $W_i = (1+0.21)W_q$ = 1.21 * (19.5~eV) = 23.6~eV, approximately~\cite{Miyajima:1974zz,Doke_2002}.} Note LAr's $W_q$ has been labeled $W_{ph}^{max}$~\cite{Doke_2002}, as it is related to the maximum possible $L_y$, when $Q_y = 0$, but this is not possible even at 0~V/cm (it does become possible as ionization density from interactions goes to $\infty$ and forces recombination in both LXe and LAr~\cite{TANAKA2001454,Doke_2002}). Drift electric field is $\mathcal{E}$ while $k_B$ is known as Birks' constant, and $A$ is the correction factor explained later. $Q_0$ is effectively the maximum possible charge, at infinite $\mathcal{E}$, defined as $E/W_i$ or $N_i$. While Birks is not the only possible parameterization (e.g., there is also the Thomas--Imel box model~\cite{ThomasAndImel,ThomImelBiller}) the focus of this work is on energy reconstruction not the various microphysics models. We thus take Birks to only be a representative example here.

Another approach, taking into account not just excitation vs.~ionization, but $e^-$-ion recombination, begins the same as for LXe~\cite{Dahl:2009nta,Sorensen_2011,Lenardo_2015}, and LAr experiments used for DM instead of neutrinos:
\vspace{-15pt}

\begin{equation}
E = N_q W_q = (N_{ph}+N_{e-}) W_q ~~\mathrm{where}~~ N_{ph} = N_{ex} + r(E,\mathcal{E}) N_i ~~\mathrm{and}~~ N_{e-} = [1-r(E,\mathcal{E})] N_i .
\label{Eqn12}
\end{equation}

\vspace{-5pt}
Here $r$ refers to the recombination probability and depends not only on energy and electric field but also the particle or interaction type, $N_{ex}$ is the number of excited atoms initially produced, and $N_i$ the number of $e^-$-ion pairs. In this case, $N_{ph}+N_{e-}$ is also $N_{ex}+N_i$. Applying the same revision to Birks' Law suggested in 2011 by Szydagis et al.~for Xe~\cite{Szydagis_2011} to Ar recasts it in terms of first principles:
\vspace{-5pt}

\begin{equation}
r(E,\mathcal{E}) = \frac{k(\mathcal{E})dE/dx}{1+k(\mathcal{E})dE/dx}~,~~ N_{e-} = (1-r)N_i \rightarrow \frac{N_{e-}}{E} = Q_y = (1-r)\frac{N_i}{E} = \frac{1-r}{W_i} = \frac{1/W_i}{1+k(\mathcal{E})dE/dx},
\label{Eqn13}
\end{equation}

with $k(\mathcal{E})=k_B/\mathcal{E}$ only one possible parameterization of recombination's E-field dependence, and in turn charge and light yields, with a more general negative power law possible~\cite{Szydagis_2011,Obodovskiy_2005}. Only the $A$ from ICARUS~\cite{AMORUSO2004275} is lacking in terms of robust justification, but it is likely a correction needed only when the secondary particle production range cut in the Geant simulation is set too high to allow for delta-ray formation down to keV-scale energies. Delta rays are lower-energy, higher-$dE/dx$ tracks with greater recombination, hence more light, at the expense of charge, explaining why $A < 1$. Increased simulation time and memory usage associated with lowering the secondary particle production range cut often lead this cut to be set too high in Geant simulations to capture this effect.

\begin{figure}[bh!]
\centering
\includegraphics[width=0.85\textwidth,clip]{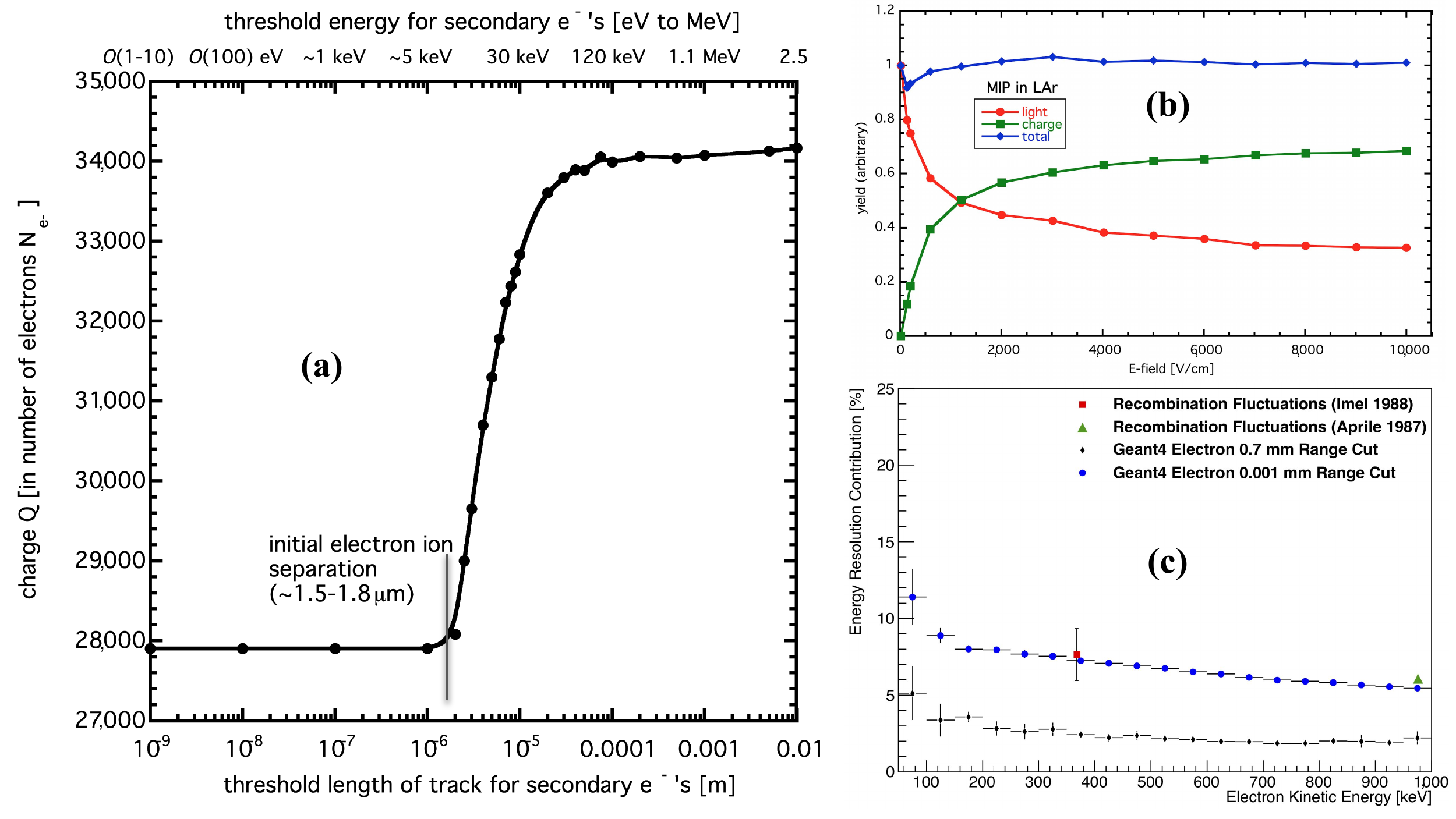}
\vspace{-0pt}
\caption{(\textbf{a}) Lack of delta rays in Geant4 altering median Q at 1~MeV in a large 500~V/cm LAr volume, more so as the threshold in track length set by the user for creation of secondary $e^-$s increases past the value from Mozumder~\cite{Mozumder_1995}. The default threshold is typically too high by $>$10 times, in length and/or $E$, in neutrino simulations~\cite{Snider_2017}. (\textbf{b}) Demonstration of S versus~Q anti-correlation from our reanalysis of Doke et al.~1988 and 2002~\cite{Doke_1988,Doke_2002}, confirmed by the recent LArIAT study~\cite{Foreman:2019dzm}. A 1~MeV ER at different E-fields, as opposed to many $E$s at one E-field (as done for a contemporary ``Doke plot''). (\textbf{c}) Noiseless charge-only resolution vs.~fixed electron $E$ using Geant4, with G4 secondary production range cuts of 1~$\mu$m (blue) and 0.7~mm (black, in LArSoft), compared to data points at isolated $E$s on recombination fluctuations with delta rays from \cite{ThomImelBiller} (red) and \cite{Aprile_1987} (green, no measurement uncertainty provided).}
\label{Fig8}
\vspace{-0pt}
\end{figure}

In Figure~\ref{Fig8}a, the number of ionization electrons produced from a 1~MeV primary electron is plotted versus both the Geant4 secondary production range cut (``length scale factor'') as well as the associated energy threshold for delta rays. The ratio between the values of the two plateaus at the left and right extremes in Q is almost exactly equal to the ICARUS best-fit value of the renormalization constant $A$ (0.800~$\pm$~0.003 in Amoruso et al.~2003~\cite{AMORUSO2004275}). The default secondary production range cut of 0.7~mm used in the LArSoft (Geant4) simulation allows for only $\sim$270+~keV delta rays to form~\cite{Snider_2017}.

\vspace{-0pt}

Figure~\ref{Fig8}b (upper right) shows that the lower plateau in the large left plot (a) in its lower left corner is closer to correct. The zero-electric-field light yield for a 1 MeV beta, or other electron, is approximately 41~photons/keV~\cite{Doke_1988} while the red line in (b) shows a reduction to $\sim$0.6 of that value by 500~V/cm, to 24.6 photons/keV. If the work function is $19.5\pm1.0$~eV, as reported also by Doke, this implies a total (S and Q summed) of $51.3^{+2.8}_{-2.5}$ quanta/keV (an S value different than 41, potentially higher, only strengthens our following arguments by lowering Q). The charge yield is total minus light, leading us to $51.3-24.6=26.7^{+2.8}_{-2.5}$~$e^-$/keV. The NEST (2012) value is thus very much within the error envelope of actual data for the lower level, but outside it for the higher one, which also never quite flattens (Figure~\ref{Fig8}a upper right). Including more relevant measurement uncertainties does not sufficiently explain this discrepancy. A similar conclusion can be reached by converting relative charge (green curve in Figure~\ref{Fig8}b) to absolute, pointing again toward the lower value of Q being the more correct one. Figure~\ref{Fig8}c demonstrates fully accounting for delta rays is also important for correctly predicting ionization-only $E$ resolution for primary electrons at these energies ($\sim$1~MeV).

The 2012--2013 version of the NEST LAr ER model was a combination of the Birks-Law and Box (Thomas--Imel) models of recombination, with LET (Linear Energy Transfer) $<$~25~(MeV$\cdot$cm$^2$)/g driven primarily by Birks, as given by Equation~(13) (with the Thomas--Imel box model still called for accompanying delta rays). There was no need for renormalization as in Equation~(11) (so that $A=1$) due to using a significantly smaller secondary production range cut in Geant4, and the Birks' constant's electric field dependence was $k(\mathcal{E}) = 0.07/\mathcal{E}^{0.85}$ ($cf.$~Amoruso's $k(\mathcal{E})=(0.0486\pm0.0006)/\mathcal{E}$). The different power on the electric field dependence, less than 1, on top of the different constant within the numerator is likely due to using Birks to model the recombination directly instead of $R=1-r$~or~$\frac{Q}{Q_0}$. In addition, it was based on higher-LET dark matter detector data and extrapolated lower. This is also nearly identical to Equation (8) from Obodovskiy's comprehensive report~\cite{Obodovskiy_2005}.

\begin{figure}[ht]
\includegraphics[width=0.95\textwidth,clip]{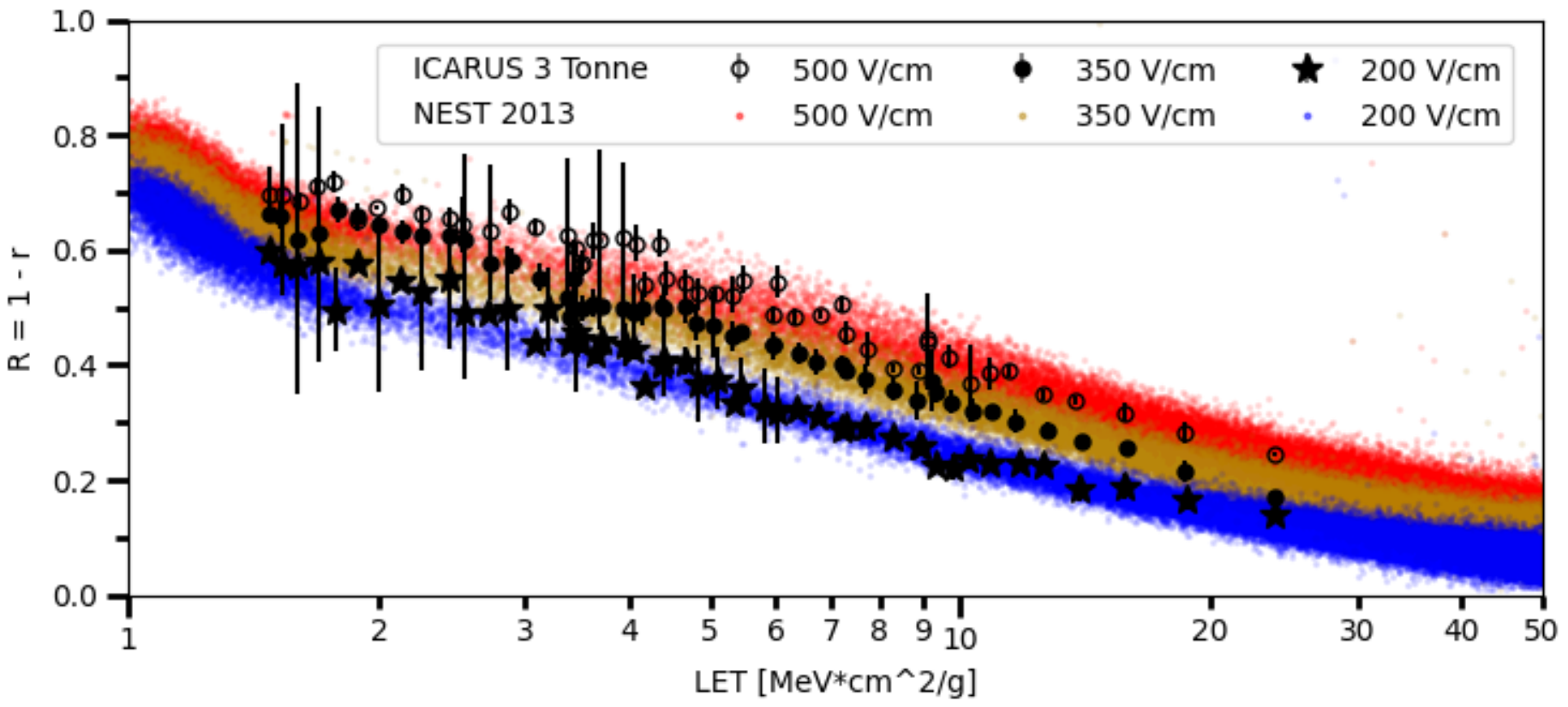}
\vspace{-5pt}
\caption{$R$ vs.~$dE/dx$ in ICARUS data (stopping muons, protons)~\cite{AMORUSO2004275} at fields 500, 350, 200~V/cm. Data in black/white, as white circle, black circle, and black star, while results of a Geant4 MC making use of the NEST 2013 LAr ER model are colored red, yellow, and blue respectively, with decreasing field; protons, muons, electrons, and their respective antiparticles, while varying in $R$ at the 5\% level, are sufficiently consistent to combine. The x-axis here is  linear energy transfer (LET), or, $dE/dx$ divided by density (NEST points extracted by individual secondary track from 1~GeV total-$E$ simulations).}
\label{Fig9}
\vspace{-0pt}
\end{figure}

One point of confusion to address is the most common quantity used for data/MC comparison in the neutrino field, the recombination factor, $R$. It is actually an ``escape'' factor for electrons from ionized atoms: $Q = (1-r)N_i \rightarrow R = 1-r = Q/N_i = Q W_i / E$, where $W_i=23.6\pm0.3$~eV~\cite{Miyajima:1974zz}. See the NEST 2013 comparison to ICARUS data (from both muons and protons mixed) in Figure~\ref{Fig9}, where a good match is observed. This is a different NEST version from the dark matter experiment (DarkSide) comparisons earlier at lower energies (higher $dE/dx$) because at time of writing the ER model for LAr in the latest NEST version has not been recast yet into a $dE/dx$ basis for a robust comparison~\cite{KateJustNote,ESTARWebSite} assuming a typical LAr density of 1.4~g/cm$^3$. Minor discrepancies are observed at low $dE/dx$, which will be addressed with planned improvements to the NEST LAr ER model in the near future.

The corrections discussed here are important not only for energy reconstruction considerations but also for background discrimination, such as for differentiating neutral pions, photons, and $e^+e^-$ pair production from single electrons which comprise the signal of interest for accelerator neutrino experiments. This particle identification can be carried out with the use of $dE/dx$ by first measuring $dQ/dx$~\cite{Acciarri_2019}. Potentially one can add in $dS/dx$ for a low-enough energy threshold, if the $g_1$ is high enough, as discussed earlier. The exclusion of delta rays common in typical Geant4 simulations for LArTPC neutrino experiments affects not only the mean yields, which can still be simulated accurately with a 20\% correction almost independently of electric field and LET, but also resolution: delta rays will degrade the energy resolution and require more complicated corrections if they are deactivated in Monte Carlo simulations (in Figure~\ref{Fig8}c). Incorrectly-modeled energy, and thus $dE/dx$, resolution can lead to potential biases between data and simulation when using reconstructed $dE/dx$ for particle identification: electron/photon discrimination in LArTPC neutrino experiments for example. 

\subsection{Liquid Argon Nuclear Recoil (Dark Matter Signal, and CEvNS)}

The final topic is low-energy (keV-scale) nuclear recoil or NR within LAr, which is important not only for dark matter in experiments such as DS~\cite{Agnes_2018_S2Only1} but also for CE$\nu$NS on collaborations like COHERENT~\cite{Akimov:2020czh}. The sum of quanta in data is again fit well by a power law, combining all global data on total yields, related to the Lindhard factor $L(E)$ from the beginning, Equation~(1).

\begin{figure}[bh!]
\centering
\includegraphics[width=0.9\textwidth,clip]{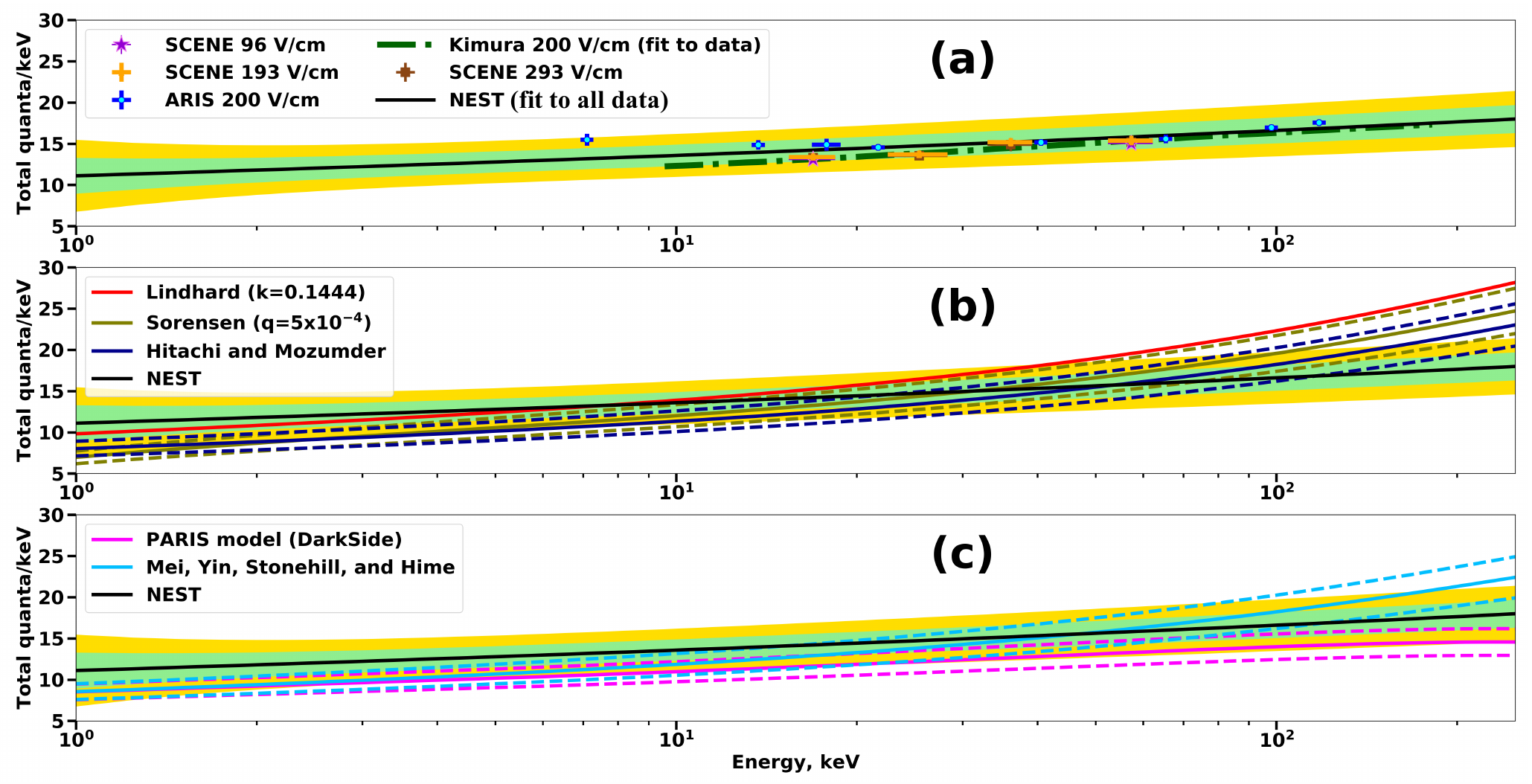}
\vspace{-0pt}
\caption{(\textbf{a}) Total number of quanta (E-field-independent) $N_q$ per keV for LAr NR: $L_y + Q_y$, or ($N_{ph} + N_{e-})/E$. Best-fit power law, used in the current NEST NR model for LAr, is $(11.10\pm1.10)E^{0.087\pm0.025}$, surprisingly similar to LXe, with 1$\sigma$/2$\sigma$ error bands in green/yellow. Kimura data~\cite{Kimura_2020} are given as a fit to the data in the original paper; SCENE and ARIS points are taken from \cite{PhysRevD.91.092007} and \cite{Agnes_2018_PRD}. (\textbf{b}) A review of models, collected from \cite{Lindhard_1963,Sorensen_2015,hitachi2019properties}, covering Lindhard/Lindhard-like approaches. Solid lines assume 45 photons/keV for the (0~V/cm) $L_y$ of ER in LAr, with  upper and lower dashes covering 50 and 40, respectively, to span the uncertainty in light, before addition with $Q_y$. (\textbf{c}) As in (\textbf{a}, \textbf{b}), NEST is repeated here with two additional model comparisons, from PARIS (used by DS~\cite{Agnes_2017}) and \cite{Mei_2008}.}
\label{Fig10}
\vspace{-0pt}
\end{figure}

The fit in Figure~\ref{Fig10} is quite similar to that from LXe earlier in Figure~\ref{Fig5} in both the base and exponent. Here we plot total quanta/keV instead of just quanta for greater clarity: in terms of quanta the power would be 1.087 ($cf.$~1.068--1.1, Xe). Data are scarce for $N_q$ in Figure~\ref{Fig10}a, as most LAr data for DM traditionally were 0~V/cm (e.g., microCLEAN, DEAP) thus lacking $N_{e-}$ to add to $N_{ph}$. $L_{eff}$ data exist from many sources, collected in~\cite{Creus_2015,Gastler_2012,Regenfus_2012,PhysRevD.88.092006} but these do not directly inform $L$, being just $L_y$.

\vspace{-0pt}

\begin{figure}[t]
\centering
\includegraphics[width=0.85\textwidth,clip]{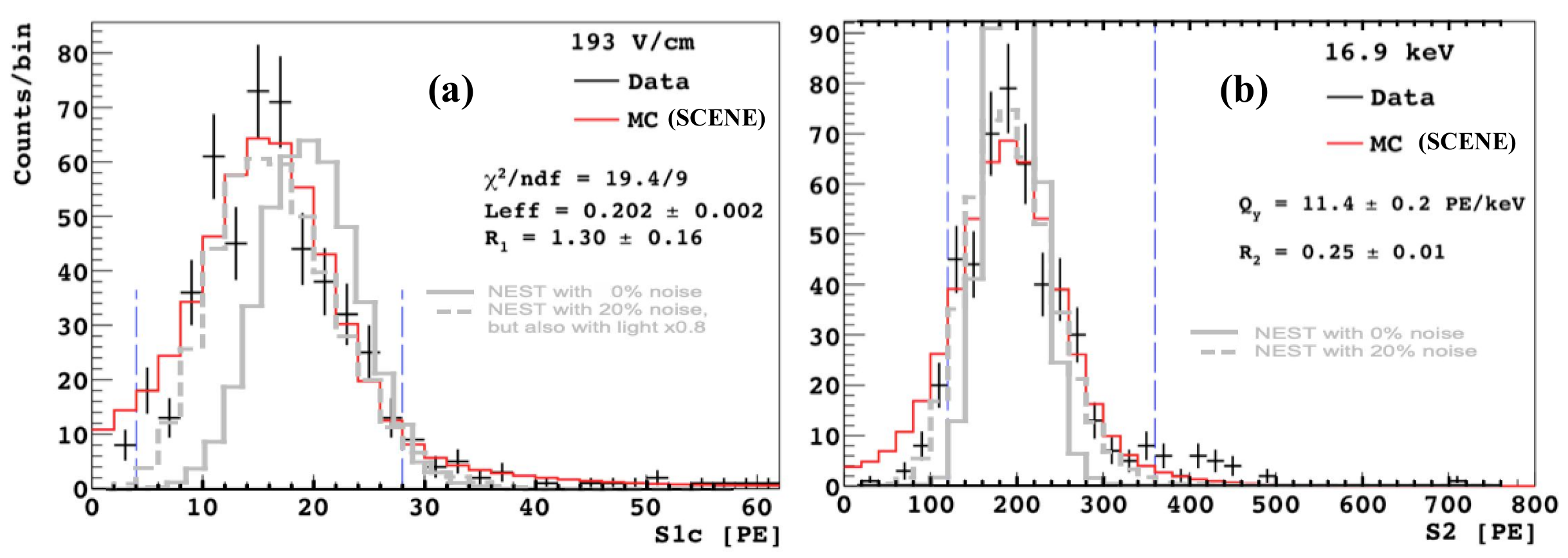}
\vspace{-2pt}
\caption{(\textbf{a}) S1 peak and (\textbf{b}) S2 peak for 16.9~keV NR in LAr from Cao et al.~\cite{PhysRevD.91.092007}. Data are in black with errors, with original MC in red (fit borders for it as vertical blue lines) and NEST overlaid in gray over original paper plots: defaults solid, while altered to match data in dash. Large noise levels needed in NEST are comparable to those assumed by SCENE ($R_1$ and $R_2$ in legends imply S1 30\%, S2 25\%).}
\label{Fig11}
\vspace{-5pt}
\end{figure}

Lindhard works surprisingly well according to Figure~\ref{Fig10}b, staying within 2$\sigma$ for the power fit to data below 50~keV. Deviation below Lindhard at high $E$ is easy to understand as bi-excitonic quenching again as per Hitachi, while (some) deviation at sub-keV is also understandable, as Lindhard's approach should break down at $<$1~keV regardless. The total number of quanta per keV is remarkably flat from 1--300~keV, around 15~quanta/keV. Breaking the total down into light and charge, a similar behavior is observed as in LXe, with $L_y$ increasing in energy, while $Q_y$ decreases~\cite{Agnes_2018_PRD, PhysRevD.91.092007}. Confusion between $L_y$ and total yield has led to past claims Lindhard does not fit well~\cite{PhysRevD.91.092007, hitachi2019properties}. Regardless of whether Lindhard fits well or does not for some energy regime, Figure~\ref{Fig10}a is indicative of a combined-$E$ scale being worthwhile (in place of S1 only) as in LXe, as argued both by DS~\cite{Pagani:2017} and LArIAT~\cite{Foreman:2019dzm}.

The final plot, Figure~\ref{Fig11}, displays the breakdown into the S1 and S2 pulse areas for one example energy coming from the same quasi-monoenergetic technique cited earlier for table-top LXe yield measurements (for both ER and NR) relying upon coincidence between a noble-element detector and a second detector, typically Ge~\cite{Plante_2011,Goetzke_2017,Agnes_2018_PRD}. For NEST MC, the means of their values for the gains have been assumed: $g_1 = 0.104 \pm 0.006$ PE/photon and $g_2 = 3.1 \pm 0.3$ PE/$e^-$. The drift electric field was 193 V/cm and sample energy 16.9~keV, selected due to being the lowest value for which the S1 and S2 histograms are displayed at an identical energy anywhere within the existing literature.

NEST seems to overestimate the $L_y$, but a larger issue is at play: contradictory data exist for NR from over the past $\sim$two decades: some experiments showed the amount of scintillation staying flat, others decreasing, and several even increasing with decreasing recoil energy~\cite{Regenfus_2012, Gastler_2012, Creus_2015}, physically possible if one adapts the logic from Bezrukov, Kahlhoefer, and Lindner from LXe on screening~\cite{Bezrukov_2011}. Those showing increase cannot be easily explained as caused by threshold bias, as several distinct sets of results agreed upon the increase~\cite{Agnes_2018_S2Only1,PhysRevD.91.092007}. Other hypotheses include different collection efficiencies, especially as wavelength shifters are generally used in LAr, and different ones~\cite{Broerman_2017}, some of which may have efficiencies higher than 100\% effectively due to creation of multiple visible photons per individual input photon in the extreme UV, though the data on the existence of this are also contradictory~\cite{Gehman_2011}. A related effect may be something in photo-sensors exposed to LAr analogous to the 2-PE effect seen in LXe. A compounding problem is the discrepancy for the zero-field $L_y$ for $^{57}$Co (or $^{83m}$Kr) used as in LXe to set $L_{eff}$. Historically, it is stated as near 40~photons/keV~\cite{Doke_1988} but more recent work, with $g_1$ known by Doke plot, suggests closer to 50, nearer the max ($1/W_q$) if $Q_y\rightarrow0$, sensible for $\mathcal{E}\rightarrow0$~\cite{Kimura_2019} (though as explained earlier, unrecombined $e^-$s remain). Systematic $L_y$ uncertainty at $\sim$20~keV may be as high as $40\%$ ($L_{eff} = 0.35$ or 0.25) making the 20\% shift ($\times$0.8) needed in Figure~\ref{Fig11}a reasonable.

At time of writing, NEST's compromise solution to the contradiction is to follow the predictions of not just Lindhard but all existing first-principles models, as well as the most recent data (ARIS) suggesting a flat-ish $L_y$ building into a mild increase with increasing NR energy, but at a higher level than the traditional $\sim$10~photons/keV (0.25 $L_{eff}$ $\times$40), better matching the claims of high $L_y$ at low energy~\cite{Agnes_2018_PRD}. (Earlier versions would switch between differing, mutually exclusive solutions.)

Figure~\ref{Fig11}b suggests an S2-only scale may ironically be more reliable for (NR-based) DM searches with LAr. Contradictions are fewer between data and different models, including NEST, even at other energies, and multi-ms $e^-$ lifetime may be easier to achieve through high-level purification~\cite{Agnes_2015,2014AIPC.1573.1169T}; however, a note of caution that this may be due to the paucity of non-zero-field data, for measuring $Q_y$, at multiple E-fields. Columnar recombination, not currently simulated, which changes the yield depending on electric field orientation, is a further complication~\cite{PhysRevD.91.092007}. The combined-$E$ scale essentially removes the effect of these and any other recombination fluctuations, which cause the variance in the original numbers of photons and electrons to be identical ``at birth'' prior to any fluctuations due to propagation and/or detection~\cite{Conti_2003,Dahl:2009nta}. Such a scale will also remove the effect of delta rays (ER) on the ultimate energy resolution achievable, if driven by anti-correlated fluctuations (recombination).

S2 is generally easier to measure, however, than the S1 is: electrons drift upward in a TPC along nearly-straight lines (slight diffusion occurs) from the liquid to the wires or to the wires followed by gas, where many photons are produced per electron. Losses due to the impurities along the drift length can be quantified simply, with an exponential. On the other hand, scintillation photons are produced in all directions and are affected by not only attenuation (not necessarily exponential, but difficult to quantify analytically) but geometric collection efficiency driven by reflection and refraction, and QE. As energy decreases, $Q_y$ increases for both NR and ER, for both LAr and LXe, at least down to the keV level before turning around, while $L_y$ appears to decrease toward 0. For this reason, NEST models are created using the total yield first, then charge yield, and light reconstructed by subtraction~in~the~code. (The same is true for LXe.)

\subsection{Liquid Argon Summaries}

In conclusion, the key points of the LAr ER section, coupled to reasonable/common detector parameter values including/especially $g_1$ and E-field $\mathcal{E}$, are:
\vspace{-0pt}

\begin{itemize}[leftmargin=*,labelsep=5.8mm]
\item A combined S1 + S2 scale continues to reconstruct ER energies best for DM/neutrino experiments, due to anti-correlation between channels, but not if $g_1$ is very low ($\ll$1\%) or $g_2$ very high (e.g., 2-phase TPC). An additional challenge is created by sitting on top of a continuous background like the beta decay of $^{39}$Ar for combined energies, but noise in Q can make S1 more favorable.
\item $dE/dx$ is more important than just $E$ at the GeV scales of greatest relevance to neutrino projects and it is most commonly reconstructed utilizing $dQ/dx$ (ignoring $dS/dx$).
\item A correction ($\sim$0.8) must be inserted into the simulation of charge yields for use in the traditional Q-only scale, lowering the Q that is output, if the delta-ray production threshold is set above the $e^-$-ion thermalization radius O(1~$\mu$m) in MC. Energy resolution may also be affected, not just mean yields, and high-energy, low-$dE/dx$ (MIP) interactions are not immune to this problem due to secondary particle production, handled with, e.g., Geant4.
\item Due to differences in delta rays and other secondaries, an analytical fit may be impossible across all particle types, leading to different recombination probabilities even if you consider only the averages versus $dE/dx$ or energy.
\item Either escape probability or recombination can be modeled as a function of the $dE/dx$ (or the LET, which includes the effects of density).
\end{itemize}

\vspace{-0pt}
Next, we summarize the key points of the LAr NR section. We note that because yields change slowly with increasing field especially for the dense tracks of NR that our S1 examples from two-phase TPCs (DS, ARIS, SCENE) should be relevant/applicable to 0~V/cm single-phase detectors as well.
\vspace{-0pt}

\begin{itemize}[leftmargin=*,labelsep=5.8mm]
\item While possible to measure for only approximately monoenergetic peaks, a summation of the few available $N_{ph}$ plus $N_{e-}$ data sets results in evidence for NR anti-correlation (akin to ERs) and modest agreement with Lindhard. This is important for both DM and CE$\nu$NS.
\item Due to uncertainty in the scintillation yield, an S2-only scale may be beneficial, but exploration of combined $E$ may still be interesting in the future (as stated above). Non-zero-field measurements are not as plentiful for charge yields as zero-field light-only ones for NR in liquid argon.
\end{itemize}

%%%%%%%%%%%%%%%%%%%%%%%%%%%%%%%%%%%%%%%%%%
\vspace{-0pt}
\section{Discussion and General Conclusions}
\vspace{-0pt}

We have reviewed mainly LXe and LAr, in dual-phase TPCs, and managed to extract insights spanning dark matter and neutrinos, proving once again the remarkable consistency obtained across numerous data sets, and the utility of NEST to probe at least the few simple reconstruction methods reviewed. From the historical perspective, it is intriguing that the application of noble elements to the DM field began with scintillation-only energy measurements, with charge primarily utilized for position reconstruction, while for neutrinos the opposite occurred: ionization-only $E$ scale, with initial scintillation used as a trigger for event activity of interest. Moving forward in both fields, it will be interesting to see how charge and light are combined together to improve $E$ measurements further than what has already been achieved.

The new insights we have gleaned from our review and meta-analyses with our own simulations, based on the global, cross-experiment framework of NEST, first for liquid xenon, include:

\vspace{-0pt}
\begin{enumerate}
\footnotesize
\item The first comparison as far as we know of the same one monoenergetic ER peak ($^{37}$Ar calibration) across S1-only, S2-only, and two versions of combined energy (standard and optimized) with both real data and NEST, with skew-Gaussians adjusting for detection efficiency and other effects. Width for S1 only was shown to be $\sim$4x worse than the best possible.
\item The only full explanation published for an optimized (weighted) combined-$E$ scale (not in a thesis or an internal report).
\item While the combined-$E$ scale has already been established as superior to S1-only in past work, we explore also an S2-only scale and show it may outperform combined energy at the keV level, but only for monoenergetic peaks and high $g_2$.
\item Demonstration that combined energies (even non-optimal) improve not just the widths and thus energy resolution, as already established in the community, but also reconstructed mean energies, and shape (i.e., symmetry or skewness).
\item Replication of measured upward bias in $E$ reconstruction with NEST, suggesting it is due to both thresholds and physics.
\item A summary of energy resolutions from $0\nu\beta\beta$ experiments, with MCs suggesting where to make improvements.
\item Clear delineation of the difference between recombination fluctuations, which affect S1 vs.~S2, and the Fano factor, that controls their total, as the literature is currently unclear on this, with one term often being used incorrectly for the other.
\item A complete comparison analogous to (1) above for the efficiency vs.~reconstructed $E$, for both ER/NR, as reconstructed by S1, S2, and both (standard and optimum combination) compared together for a simple continuous spectrum (box/WIMP).
\item The most complete compilation to date of NR $Q_y+L_y$, showing also a Lindhard-like power law matches the total yield.
\normalsize
\end{enumerate}

\vspace{+10pt}
For liquid argon, items of interest presented here that were new, to the best of our~knowledge, are:
\vspace{-5pt}

\begin{enumerate}
\footnotesize
\item The first comparisons of NEST performed for LAr, in plots for dark matter at low energies and neutrino physics for high $E$s (vs.~$dE/dx$) never presented before, demonstrating a new understanding of mean yields and widths, requiring G4.
\item An exhaustive simulated table relevant to neutrino physics that goes beyond existing data, and predicts a significant improvement in $E$ reconstruction at an energy (1 MeV, low LET) still relevant to neutrinos, for sufficiently high $g_1$.
\item A demonstration that anti-correlation was ``hiding'' in seminal work by Doke et al.~with an explicit reanalysis of the original paper showing photons and electrons sum to a constant for a MIP in LAr as a function of E-field at fixed $E$.
\item A clarification of confusing/conflicting definitions of work function, recombination probability, and charge yield.
\item Quantitative proof confirming the hypothesis ICARUS' data required a correction specifically for not having delta rays activated in simulation, plus the first evidence not just mean reconstruction of charge is affected but also the width.
\item A comprehensive compilation of all existing data and models for NR in terms of total yield not just light, beyond 0~V/cm.
\normalsize
\end{enumerate}

%%%%%%%%%%%%%%%%%%%%%%%%%%%%%%%%%%%%%%%%%%

\funding{The research of Levy and Szydagis at the University at Albany SUNY (the State University of New York) was funded by the United States Department of Energy (DOE) under grant number \textbf{DE-SC0015535}. Mooney and his students, Alex Flesher and Justin Mueller, were funded through university startup funding. Kozlova was funded through the Russian Science Foundation (contract number 18-12-00135) and Russian Foundation for Basic Research (projs.~20-02-00670a).}

%%%%%%%%%%%%%%%%%%%%%%%%%%%%%%%%%%%%%%%%%%
\acknowledgments{The authors wish to thank these members of the LZ and/or LUX collaborations who reviewed a draft of the Xe section informally and provided helpful comments: Henrique Ara\'ujo of Imperial College London, Andrew Stevens of Oxford University, and Chamkaur Ghag of University College London. We also acknowledge Dan McKinsey along with all his students and postdocs who worked on PIXeY, with whom we have had very fruitful discussions over the years. Lastly, Szydagis thanks Jason Sokaris and Steve Cifarelli for early work on the optimized energy scale, as well as Dmitri Akimov of MEPhI, for allowing his advisee E.~Kozlova to work on NEST, and on this paper.}

\conflictsofinterest{The authors declare that the author list of this paper includes members of the LZ, LUX, nEXO, DUNE, MicroBooNE, SBN, CENNS-10, COHERENT, and RED collaborations.
}
%%%%%%%%%%%%%%%%%%%%%%%%%%%%%%%%%%%%%%%%%%
\reftitle{References}

\end{document}